\documentclass[aoas,MSNbibl,nameyear,dvips]{arximspdf}
\usepackage{ifthen,mathbh}
\usepackage{graphicx}

\doi{10.1214/10-AOAS402}
\volume{5}
\issue{4}
\pubyear{2011}
\firstpage{2326}
\lastpage{2358}

\makeatletter
\newcommand{\paragraphm}[1]{\textit{#1.}}
\newcommand{\eqref}[1]{(\ref{#1})}

\newcommand{\locweightm}{\mathbf{w}_{\nsites}}
\newcommand{\spweightm}{\mathbf{w}_{\nspecies}}

\newcommand{\myvec}[1][]{
    \def\ArgI{{#1}}
    \vecRelay
    }
\newcommand{\vecRelay}[2][]{
    \ifthenelse{\equal{#1}{}}
        {{\mathbf{#2}_{\ArgI}}}
        {{\mathbf{#2}_{\ArgI {(#1)}}}}
}
\newcommand{\vc}{\myvec}
\newcommand{\mat}[2][]{ 
    \ifthenelse{\equal{#1}{}}
        {{\mathbf{#2}}}
        {
        \ifthenelse{\equal{#1}{plain}}
            {{\mathbf{#2}}}
            {{\mathbf{#2}_{#1}}}
        }
}
\newcommand{\newx}{\hat{x}}
\newcommand{\newX}{\widehat{X}}
\newcommand{\one}[1][]{\vc[#1]{1}}
\newcommand{\id}[1][]{\mat{I}_{#1}}

\newcommand{\cov}{\operatorname{cov}}
\newcommand{\var}{\operatorname{var}}

\newcommand{\mhat}[1]{\mat{\widehat{#1}}}
\newcommand{\mtilde}[1]{\mat{\widetilde{#1}}}

\newcommand{\mc}[1]{{\mathcal{#1}}}
\newcommand{\half}{\frac{1}{2}}
\newcommand{\R}[1]{\mathbb{R}^{#1}}
\newcommand{\ind}[1]{\mathbh{1}\{#1\}}

\newcommand{\nrow}{{n}} 
\newcommand{\ncol}{{p}}
\newcommand{\colmetric}[1][\nrow]{\mat{Q}_{#1}}
\newcommand{\rowmetric}[1][\ncol]{\mat{Q}_{#1}}

\newcommand{\rowmetricLetter}{Q}
\newcommand{\matcoltrans}{\mat{B}}
\newcommand{\rowcov}[1][]{\mat[#1]{\bolds{\Psi}}}
\newcommand{\proj}[1][]{\mat{P}_{#1}}

\newcommand{\treedist}[2]{{d_\mathcal{T}(#1,#2)}} 
\newcommand{\troot}{{\mathcal{R}}} 
\newcommand{\tcov}[1][]{\mat[#1]{\bolds{\Sigma}}}

\newcommand{\locweight}[1][]{
        \ifthenelse{\equal{#1}{}}
                {\myvec[  _{\nsites}]{w}}{\myvec[ _{\nsites}][{#1}]{w}}} 
\newcommand{\spweight}[1][]{
        \ifthenelse{\equal{#1}{}}
                {\myvec[ _{ \nspecies}]{w}}{\myvec[ _{\nspecies}][{#1}]{w}}} 
\newcommand{\wloc}{\mathbf{D}_{\locweightm}} 
\newcommand{\wsp}{\mathbf{D}_{\spweightm}} 
\newcommand{\locprofile}[2][]{\vc[#2][#1]{x}}
\newcommand{\locmat}[1][]{\mat[#1]{X}}
\newcommand{\locbar}{\vc{\bar{x}}} 
\newcommand{\tildlocmat}[1][]{{\widetilde{\locmat[#1]}}}

\newcommand{\ab}[1][]{\mat[#1]{A}}
\newcommand{\atotal}{N}

\newcommand{\divgs}{{H_{\mathrm{GS}}}}
\newcommand{\divsh}{{H_{\mathrm{Sh}}}}

\newcommand{\dist}[1][]{\mat[#1]{\bolds\delta}}
\newcommand{\dpx}{\mat{Z}}
\newcommand{\dpy}{\mat{Y}}
\newcommand{\sitecoord}{\mat{L}}
\newcommand{\speciescoord}{\mat{K}}

\newcommand{\nspecies}{S}
\newcommand{\nsites}{{L}}

\newcommand{\similar}{\mat{S}_{\vc{v}}}
\newcommand{\adj}{A}
\newcommand{\lap}{\mat{L}}
\newcommand{\kernel}[1][]{\mat{K}_{#1}}

\makeatother

\begin{document}
\begin{frontmatter}

\title{Analysis of a data matrix and a graph: Metagenomic data and the
phylogenetic tree}
\runtitle{Metagenomic data and the phylogenetic tree}

\begin{aug}
\author{\fnms{Elizabeth} \snm{Purdom}\corref{}\ead[label=e1]{epurdom@stat.berkeley.edu}\thanksref{t1}}
\thankstext{t1}{Supported in part by a NSF Post-doctoral Fellowship in
Biological Informatics and by NSF Grant 02-41246.}
\runauthor{E. Purdom}
\affiliation{University of California, Berkeley}
\address{Department of Statistics\\
University of California at Berkeley\\
367 Evans Hall \#3860\\
Berkeley, California 94720-3860\\USA\\
\printead{e1}} 
\end{aug}

\received{\smonth{11} \syear{2009}}
\revised{\smonth{7} \syear{2010}}

%
\begin{abstract}
In biological experiments researchers often have information in the
form of a graph that supplements observed numerical data. Incorporating
the knowledge contained in these graphs into an analysis of the
numerical data is an important and nontrivial task.
We look at the example of metagenomic data---data from a genomic survey
of the abundance of different species of bacteria in a sample. Here,
the graph of interest is a phylogenetic tree depicting the interspecies
relationships among the bacteria species. We illustrate that analysis
of the data in a nonstandard inner-product space effectively uses this
additional graphical information and produces more meaningful results.
\end{abstract}

%

\end{frontmatter}

\section{Introduction}\label{sec:Introduction}
Relationships among either observations or variables are often
conveniently summarized by a graph. Incorporating this outside
information into the analysis of numerical data is of increasing
interest, particularly in biology where many known properties of genes
and proteins are described by complicated networks. A common situation
is to have numerical data from an experiment which is of primary
interest and also additional knowledge in the form of a graph relating
our observations or variables from the experiment. We would like to
incorporate the information in the graph with our analysis of the
experimental data. By including the graphical information directly in
our analysis, we constrain the space of possible solutions to those
that are relevant from the point of view of the known information.

The specific type of graph which we consider here is a phylogenetic
tree. A~phylogenetic tree is a ubiquitous graph in biology that
describes the evolutionary relationship between a set of species. We
are motivated to consider this graph by our work with \citet{myscience}
analyzing differences in bacterial composition based on a genomic
inventory of different samples. Such ``metagenomic'' studies are a
popular technique for measuring bacterial content. As we argue below,
using the phylogenetic information regarding the discovered bacteria is
key in creating a meaningful analysis---particularly because of the
small sample size relative to the number of bacteria found.

There are numerous different strategies for using graphical
information, such as Bayesian networks and differential equation
modeling; they require varying degrees of specificity in the graphical
information. We focus here on a technique that is simple to implement
and uses the graph to define a nonstandard inner-product space in $\R
{p}$ to perform the analysis of the numerical data.

The layout of the paper is as follows. First we will introduce the
motivating example of bacterial composition in more detail and will
return to the example at the end to demonstrate the techniques on the
bacterial data. We review how PCA can be succinctly reformulated for
nonstandard inner-products and its development for ecological studies
of species abundance, a~reformulation we will call generalized PCA
(gPCA). The rest of the paper delves further into the implications of
incorporating outside graphical information through the use of such a
metric space. In particular, we give an appropriate metric for a
phylogenetic tree and evaluate the implications of that choice in the
final data analysis. Throughout, we focus on the example of the
phylogenetic tree and metagenomic data to illustrate the concepts.
However, the same basic approach can be useful in including nonstandard
forms of knowledge---other types of graphical information in particular.

\paragraphm{Notation} In all that follows, we will use boldface type to
indicate vectors and matrices and parenthetical subscripts to indicate
elements of vectors and matrices. Therefore, the $j$th component of a
vector $\vc[i]{x}$ will be given as~$\vc[i][j]{x}$ and the $i,j$
element of a matrix $\mat{A}$ will be given as $\mat[(ij)]{A}$.

\section{Motivating example}\label{sec:Data}
In \citet{myscience} the broad goal was to describe the kinds of
bacteria found in the intestinal tract and compare the bacterial
communities found in different people. To that end, each of the three
patients in the study had biopsies taken at six locations in his/her
colon in addition to providing a stool sample. Each of these seven
samples (per patient) was then subjected to genomic techniques to try
to quantify the different types of bacteria as well as their abundance.

Traditional techniques for identifying bacteria require growing the
bacteria in a culture and then classifying the bacteria as a species
based on any observable characteristics as well as the nutrients needed
for it to grow. This gives only limited ability to assess the presence
of different types of bacteria. The increased ease of DNA sequencing
has led researchers to classify bacteria by genomic information
(``metagenomics''). We focus here on the results of sequencing a
specific gene (16S rDNA) found in bacteria. A random selection of all
the copies of the gene present in the sample are sequenced. Ideally,
each version of the gene could be uniquely identified as coming from a
specific bacteria and the abundance of the different gene versions
would give an estimate of the abundance of each bacteria. In reality,
we do not have a direct link between a gene version and its originating
bacteria, but only an estimate of it, as we explain more fully below.

\begin{figure}

\includegraphics{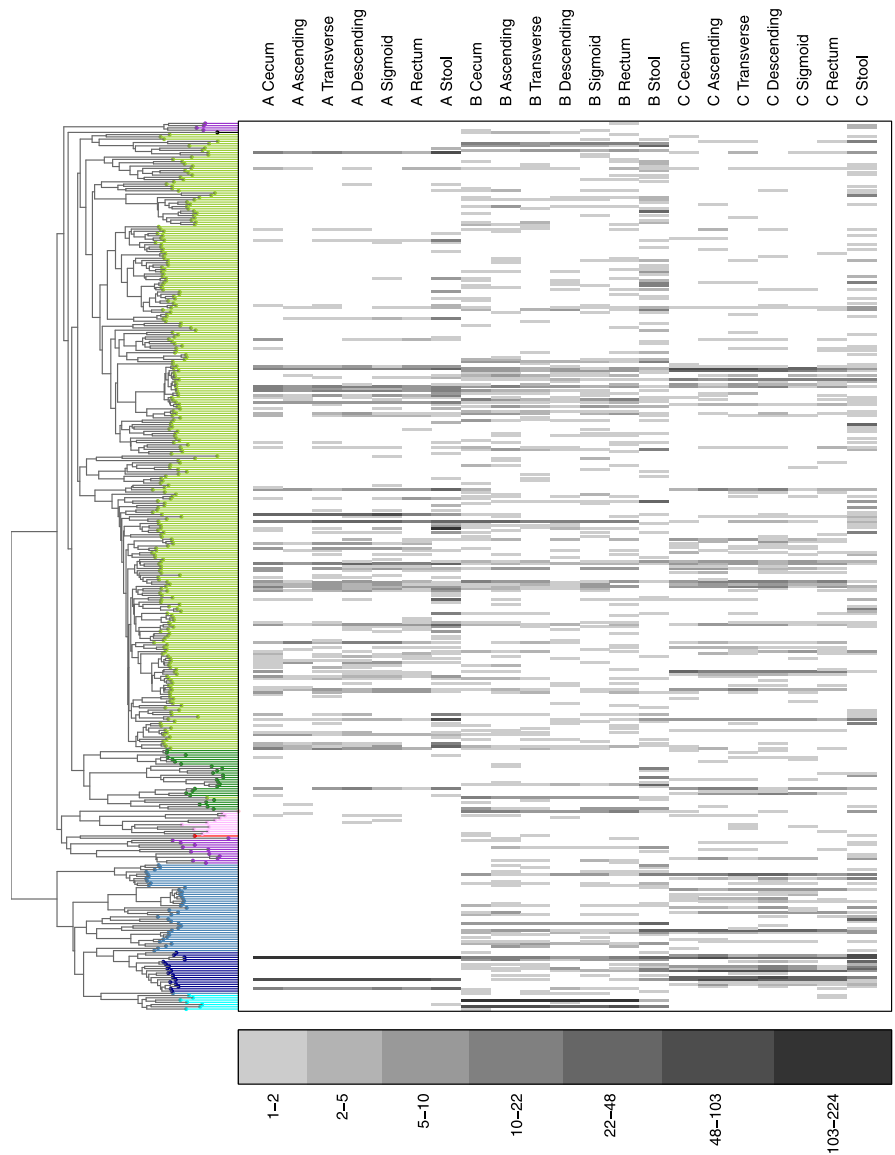}

\caption{Depiction of the abundance matrix
from  Eckburg et~al. (\protect\citeyear{myscience}). Columns indicate samples, grouped by patient,
and rows correspond to different phylotypes. The grey scale indicates
the level of abundance on a log scale (see legend for conversion to
original abundances). The colors on the phylogenetic tree indicate
phylum, as in Eckburg et~al. (\protect\citeyear{myscience}), but with a different choice of colors:
blue---\textit{Bacteriodetes}, green---\textit{Firmicutes}, purple---various \textit
{Proteobacteria}, pink---\textit{Verrucomicrobia}. We additionally colored
two portions of the \textit{Bacteriodetes} phylum (blue) separately:
roughly identifiable as \textit{Prevotallae} and \textit{B. vulgatus}, they
are colored lightest blue and darkest blue, respectively. Also, we
colored the \textit{Firmicutes} (green) with two different shades for
\textit{B. Mollicutes} and \textit{Clostridia} (dark green and light green,
respectively).}
\label{fig:intesData}
\end{figure}

Bacteria species also share an evolutionary history which might affect
their biological role in the sample. We summarize the evolutionary
relationship by a phylogenetic tree that describes the evolutionary
history of the bacterial species. We visualize both the phylogenetic
tree relating the bacterial species and their numerical abundance in
Figure \ref{fig:intesData}. There is a great deal of sparsity in the
data; many species are present in low numbers and in only a few
samples. At the same time, there are some highly abundant species found
at high levels in most samples. From this visual inspection, we can
also see the importance of jointly considering both aspects of the data---entire
regions of the phylogenetic tree appear dissimilar between
the patients, such as the \textit{Bacteriodetes} phylum (colored shades
of blue) where patient A has much less abundance across all of his/her
samples than the other two patients.

Given the large number of species (395) as compared to the number of
samples (21), we could reorder the species and find other sets of
species that are also very different across the patients. However, the
clusters defined by the phylogenetic tree provide biological
information regarding the relationships among the species that is
separate from the numerical abundances. Patterns of sparsity or
differences among the patients following the clusters in the tree are
generally of greater interest than an arbitrary grouping since there is
known biological meaning to the groupings. The additional information
found by using the phylogenetic similarities can serve as a check on
the kind of relationships among the species that we are interested in.
This will be particularly important since we have so many more species
than samples. Focusing the analysis to follow the structure of the tree
will allow for more meaningful results.

This study was exploratory. It was the first sequence-based analysis of
the bacterial composition of the colon that compared between
individuals and/or locations of the sample (many genomic experiments of
this type either sampled only one patient or pooled patients together).
The list of phylotypes found and their relationship to known bacterial
taxa was biologically informative. In addition to creating an
inventory, the goals of the experiment were to better describe the
bacteria communities and their differences along the intestinal tract
or between patients. With the small sample size, the analysis cannot
extrapolate to the population in general but can only focus on
describing the patients observed.\vfill\eject

\subsection{Effect of imperfect species definition}

In practice, we cannot identify a bacterial species from the DNA
sequence. Instead the sequences are themselves used to \textit{define} the
species, based on the sequence similarity of different copies---for
example, the rule in \citet{myscience} for grouping sequences into one
``species'' required all pairs in the group to have a minimum of 99\%
sequence similarity. For this reason, the term ``phylotype'' is used
instead of species to indicate that these are merely proxies for the
true species distinctions. A phylogenetic tree for the phylotypes was
built using maximum likelihood estimation of the tree [\citet
{Felsenstein1981}]. Specifically, the tree was built using a
representative instance of the 16S rDNA sequence from each phylotype,
generally a consensus sequence of those sequences classified into that
phylotype.

The possible effect of using an arbitrary cutoff for defining
phylotypes is seen in Figure \ref{fig:intesData}, where the length of
the tree branch reflects the similarity between the species. Some
phylotypes clearly form tight bunches of very similar phylotypes,
particularly in the \textit{Clostridia} family of the \textit{Firmicutes}
phylum (light green). If we had changed the cutoff for defining
phylotypes, we could imagine these groups collapsing into a few
distinct phylotypes. Therefore, we need to be careful to have an
analysis that is robust to such small changes and does not count each
phylotype as equally important.

The relationship between DNA sequences can be summarized in different
ways, such as its similarity to other sequences, the phylotype to which
it has been assigned, or its location in a phylogenetic tree built
between different sequences. The analysis discussed in detail here will
reduce the sequence data to the phylotype-level, ignoring the
individual sequence data: each of the $\atotal=11\mbox{,}831$ observations (or
sequenced strands of DNA) belongs to one of $\nspecies=395$ phylotypes
(or species) and one of the $\nsites=21$ locations.

\section{Incorporation of additional information via
inner-products}\label{sec:generalApproach}

For observed data $\vc[i]{x}\in\R{\ncol}$ we propose to use nonstandard
inner-products or metrics in analyzing the data. We argue that this is
a simple way to include complicated outside information, such as
graphical information, in the analysis of high-dimensional data.

By nonstandard inner-products, we specifically mean an inner-product
between two observations $i$ and $j$ given by $\langle\myvec[i]{x},\myvec
[j]{x}\rangle_{\rowmetric[]}=\myvec[i]{x}^T\rowmetric[]\myvec[j]{y}$.
Since $\rowmetric[]$ also defines a metric based on $\|\vc[i]{x}-\vc
[j]{x}\|_{\rowmetric[]}$, we may at times refer to $\rowmetric[]$ as a
metric. For any inner-product $\langle\cdot,\cdot\rangle$ and a fixed
set of $n$ vectors $\vc[i]{x}$, there exists a matrix $\rowmetric[]$ so
that $\langle\vc[i]{x},\vc[j]{x}\rangle=\vc[i]{x}^T\rowmetric[]\vc
[j]{x}$, so this is a quite general definition. A common example of
such an inner-product is the Mahalanobis distance, where $\rowmetric[]$
is chosen as the inverse covariance matrix of the observed random
vectors [see \citet{DeMaesschalck2000p418}]. In this case, the choice
of $\rowmetric[]=\mat{\rowcov}^{-1}$, where $\mat{\rowcov}$ is the
covariance matrix of the observed variables, removes the correlation
among the variables, also known as ``sphering'' the data. This is the
most common choice of a nontrivial $\rowmetric$ and is used, for
example, in discriminant analysis for classification problems.

The choice of an appropriate metric $\rowmetric[]$, however, can also
be a method for including outside information. In particular, assume
that the additional information, such as the phylogenetic tree, is such
that one can model the covariance structure ${\bolds{\Sigma}}$ for the
variables that this information would imply. The resulting covariance
matrix, $\bolds{\Sigma}$, is not the covariance for the observed
variables in our data---which is the result of a much more complicated
relationship between the graph and the data---but rather what would be
expected if the data was completely created by this outside process. In
order to evaluate the data so as to give priority to relationships in
the phylogenetic tree, we propose using the metric $\rowmetric=\bolds
{\Sigma}$ for the variable space. Performed in this space, the analysis
focuses on the aspects of the data variables most congruent with the
$\bolds{\Sigma}$.

Because most multivariate techniques are based on inner-products, they
are easily generalized to a more general inner-product space. We will
focus on PCA using $\rowmetric[]$, a technique known as generalized PCA
(gPCA) or the duality principle [\citet
{Escoufier1987}; \citet{Holmes2006}; \citet{Dray2007}]; \citet{Jolliffe2002} gives a more
in depth overview of gPCA, connecting gPCA with other techniques. We
give a short review of gPCA before we discuss more fully the
interpretation of this strategy. Other multivariate methods have been
similarly extended and would be also relevant for incorporation of
outside information.

\subsection{Generalized PCA}\label{sec:gPCA}
Quite generally, gPCA is an ordination procedure, that is, each
observed data point $\vc{x}\in\R{\ncol}$ is transformed to new,
lower-dimensional data coordinates given by $\vc{\newx}\in\R{k}$ which
is a linear transformation of the original coordinates: $\vc{\newx}=\mat
{Z}^T\vc{x}$ for some matrix $\mat{Z}\in\R{\ncol\times k}.$ Most
multivariate techniques are ordination procedures, common examples
being PCA, Canonical Correlation Analysis and Correspondence Analysis.
The differences lie in the choice of the linear transformation ($\mat
{Z}$), which is chosen based on the desired properties of the new,
lower-dimensional vector $\vc{\newx}$. The most familiar example is
standard PCA which seeks successive vectors $\vc[j]{a}\in\R{p}$ so that
the resulting $j$th coordinate, $\vc[][j]{\newx}=\langle \vc{x},\vc
[j]{a} \rangle,$ has the largest variance, subject to being independent
of previous coordinates $\vc[][1]{x},\ldots ,\vc[][j-1]{x}$; the final
transformation matrix is $\mat{Z}=\mat{A}_{k}=(\vc[1]{a}\cdots\vc[k]{a})$.

The ordination procedure of generalized PCA (gPCA) is a generalization
of PCA in that it assumes an alternative inner-product for the data
vectors~$\vc{x}$. We assume an observed random variable $\vc{x}$ lies
in $\R{p}$ with a known inner-product defined by $\rowmetric\in\R{\ncol
\times\ncol}$. Then in analogy with standard principal components, gPCA
can be developed from the perspective of finding the vector $\myvec{a}$
that maximizes the population quantity,
$\var(\langle\myvec{a}, \vc{x}\rangle_{\rowmetric}),$
with~$\vc{a}$ constrained to have unit $\rowmetric$-norm and successive
$\vc[j]{a}$ constrained to be $\rowmetric$-orthogonal to the preceding
$\vc[j]{a}$,
\[
\| \vc[j]{a}\|_{\rowmetric}=1  \quad \mbox{and} \quad  \mat{A}_{k}^T\rowmetric\mat
{A}_{k}=\mat{I}_k,
\]
where, again, $\mat{A}_{k}\in\R{\ncol\times k}$ is the matrix with
columns $\vc[j]{a}$. The new coordinates for $\vc{x}$ are then given by
$\vc{\newx}=\mat{A}_{k}^T\rowmetric\vc{x}$ (so $\mat{Z}=\mat
{A}_{k}^T\rowmetric$ in the notation given above). As in PCA, the
$\myvec[j]{a}$ will be eigenvectors, but now of the matrix $\rowcov
\rowmetric$ where $\rowcov$ is the covariance matrix of $\vc{x}$.
The matrix $\rowcov\rowmetric$ is not symmetric, but because $\rowmetric
$ is full rank, this is a well defined, positive definite generalized
eigenequation, and the eigenvectors of $\rowcov\rowmetric$ can be
chosen to be a $\rowmetric$-orthogonal set of vectors [see \citet{GolubBook}].

Just as in PCA, there are multiple developments that result in the same
ordination procedure. For example, gPCA provides the best
$k$-dimensional approximation to the inter-point similarities when the
similarities are calculated in the appropriate metric space. In
particular, we could note that if distances between observation $i$ and
$j$ are given by
\[
d(i,j)=(\vc[i]{x}-\vc[j]{x})^T \rowmetric(\vc[i]{x}-\vc[j]{x}),
\]
then gPCA is equivalent to Multidimensional Scaling (MDS) of the $\nrow
$ observations based on these distances. Similarly, for any $\rowmetric
$, there exists a~(non\-unique) matrix $\mat{C}$ so that $\rowmetric=\mat
{C}\mat{C}^T$, which means gPCA of $\vc{x}$ based on $\rowmetric$ is
equivalent to first transforming the vector $\vc{x}$ by $\mat{C}$ and
then performing PCA on the resulting vector $\mat{C}^T\vc{x}$.

\paragraphm{Metric for the columns} As an analysis of a $\nrow\times
\ncol$ data matrix $\mat{X}$, the above presentation only considered a
metric for the space of row vectors (observations) of $\mat{X}$. There
can also be a relevant metric for comparison of the variables, a simple
example being when there are weights assigned to the observations.
Generalized PCA goes beyond the description given so far and allows
also for a metric $\colmetric\in\R{\nrow\times\nrow}$ for the space of
the column vectors of $\mat{X}$. These combinations of choices are
generally abbreviated as the triplet $(\mat{X},\rowmetric,\colmetric)$
[see \citet{Escoufier1987} for a more general explanation of the role
of two separate metrics when viewing $\mat{X}$ as an operator
simultaneously in~$\R{\ncol}$ and $\R{\nrow}$]. We note that in many
cases either $\colmetric$ or $\rowmetric$ are chosen to be diagonal, in
which case they simplify to weights on the observations or variables,
respectively.

Returning to the population development above, the inclusion of a
metric for the columns of $\mat{X}$ is incorporated in the estimation
of $\rowcov.$ In order to maximize the quantity $\var(\langle\myvec{a},
\vc{x}\rangle_{\rowmetric})$, we must estimate $\rowcov$ from our data
matrix $\mat{X}$; we include the metric $\colmetric$ for the columns in
our estimate so that $\mhat{\rowcov}=\mat{X}^T\colmetric\mat{X}$. Then
our estimates of $\vc[j]{a}$ are given by the eigenvectors of $\mat
{X}^T\colmetric\mat{X}\rowmetric$. A geometric development that
includes the metric $\colmetric$ for the columns shows that gPCA best
preserves the total inner-point similarities of the data matrix $\mat
{X}$ when using a measure of inner-point similarities incorporating the
row and column matrix known as the inertia (see Appendix \ref
{app:decomposeDiv}). In addition to the geometric view of gPCA, \citet
{Jolliffe2002} notes that gPCA with $\colmetric$ a diagonal matrix
provides the maximum likelihood estimates of the fixed effects version
of a factor model,
\[
\vc{x}=\mat{A}\vc{z}+\vc{\bolds\varepsilon},
\]
where $\vc{\bolds\varepsilon}\sim N(\vc{0},\sigma^2\colmetric^{-1}\rowmetric^{-1}).$

\paragraphm{Connection between analysis of the rows and columns}
In some data settings either the rows or the columns can be
meaningfully considered as the observations, such as analysis of large
contingency tables that are our motivating example. Furthermore, the
importance of the different variables in describing a low-dimensional
representation of the observations is a common part of PCA. A gPCA of
the \textit{columns} of $\mat{X}$, also reduced to $k$ dimensions, is
technically the gPCA of triplet $(\mat{Y}=\mat{X}^T,\colmetric
,\rowmetric)$ and results in new coordinates for the columns given by
$\mat{\hat{Y}}=\mat{Y} \mat{Q}_{\nrow}\mat{B}_k\in\R{p\times k}$.

Again in analogy to PCA, a generalized form of the SVD of $\mat{X}$
yields the solutions to gPCA on both the columns or the rows
simultaneously. If the rank of $\mat{X}=r$, we can write $\mat{X}=\mat
{\matcoltrans}\bolds{\Lambda}^{1/2}\mat{A}^{T}$, where $\mat{A}\in\R{\ncol
\times r}$ and $\mat{B}\in\R{\nrow\times r}$, and the columns of $\mat
{\matcoltrans}$ are $\colmetric$-orthogonal and the columns of $\mat
{A}$ are $\rowmetric$-orthogonal. Then $\mat{\matcoltrans}$ gives the
solutions to the gPCA of the columns as observations, while $\mat{A}$
gives the solutions to the gPCA of the rows as observations. The
corresponding eigenequations are
\begin{eqnarray*}
&&\mat{X}^T\colmetric\mat{X}\rowmetric\mat{A}=\mat{A}\bolds{\Lambda},\\
&&\mat{X}\rowmetric\mat{X}^T\colmetric\mat{\matcoltrans}=\mat
{\matcoltrans}\bolds{\Lambda},
\end{eqnarray*}
and for any choice of $k$, $\matcoltrans_{k}=\mat{X}\rowmetric\mat
{A}_{k}\bolds{\Lambda}_{k}^{-1/2}$, where $(\cdot)_{k}$ refers to the
matrix with the first $k$ columns or diagonal elements, as appropriate.

This means the new coordinates from a gPCA of the rows can be
completely determined by the new coordinates from a gPCA of the columns
of the data matrix. Let $\vc{\newx}\in\R{k}$ be the new coordinates for
a vector $\vc{x}\in\R{p}$ based on the gPCA of the rows of $\mat{X}$.
The new coordinates are given as
\[
\vc{\newx}^T=\vc{x}^T\rowmetric\mat{\hat{Y}}\bolds{\Lambda}_{k}^{-1/2}.
\]
Put another way, the value of the $j$ new coordinates of $\vc{\newx}$
is given by
\[
\vc[][j]{\newx}=\langle\vc{x}, \vc[j]{\chi}\rangle_{\rowmetric},
\]
where $\vc[j]{\chi}$ is the $j$th column of $\mat{\hat{Y}}\bolds{\Lambda
}_{k}^{-1/2}$, that is, the column of $\mat{\hat{Y}}$ normalized to
have standard deviation one. Thus, the $j$th coordinate of $\vc{\newx}$
is a measure of the similarity of $\vc{x}$ with the $j$th variable
defining the reduced space of the columns.

\subsection{Interpretation of nonstandard metrics}\label{sec:harmonic}
Using a metric for $\R{\ncol}$ has an obvious rationale when the metric
is a diagonal, implying different weights for different variables, or
when the metric is $\mat{\rowcov}^{-1}$ where $\rowcov$ is the
covariance of the variables (Mahalanobis distance). However, it is not
immediately clear why a particular matrix $\rowmetric$, such as
$\rowmetric=\bolds{\Sigma}$ as we propose above, would improve a given
data analysis. One intuitive rationale for this comes from thinking of
the metric as defining a harmonic analysis of the data in the direction
of the eigenvectors of $\rowmetric$. This is the perspective of \citet
{Rapaport2007} in their proposal for the particular case of general
graphs (see Section \ref{sec:generalGraphs}).

Outside information, such as our phylogeny, when represented by $\bolds
{\Sigma}$ also defines a basis given by the eigenvectors $\vc[j]{v}$ of
$\bolds{\Sigma}$. The eigenvectors decompose our overall covariance into
hopefully informative directions with regards to our outside structure,
and the $\vc[j]{v}$ can be ordered based on their overall contribution
to $\bolds{\Sigma}$ based on the eigenvalues $\lambda_j$. The directions
given by the~$\vc[j]{v}$ can be weighted in different ways to create a
family of metrics, with each choice of weighting system emphasizing
different directions.

More precisely, suppose $\bolds{\Sigma}$ has an eigendecomposition given
by $\mat{V}\bolds{\Lambda}\mat{V}^T$; $\mat{V}$ is a $\ncol\times\ncol$
matrix with columns $\vc[j]{v}$ consisting of the eigenvectors of $\bolds
{\Sigma}$, and~$\bolds{\Lambda}$ is a diagonal matrix of eigenvalues
$\lambda_j$. The vectors $\vc[j]{v}$ form a basis for $\R{p}$ and,
therefore, a data vector $\vc{x}$ can be written as
\[
\vc{x}=\sum_j \langle\vc[j]{v},\vc{x}\rangle \vc[j]{v}=\mat{V}\vc{\breve{x}},
\]
where $\vc[][j]{\breve{x}}=\vc[j]{v}^T\vc{x}$ gives the magnitude of
$\vc{x}$ in the direction of the eigenvectors of~$\bolds{\Sigma}$.

This decomposition of $\vc{x}$ into its contributions due to the
directions given by $\vc[j]{v}$ creates no loss of information, being
only a change of basis. But we can transform the original $\vc{x}$ by
giving weights $\vc[][j]{w}$ to different directions in order to give
more emphasis to the features that $\vc[j]{v}$ represents, in which
case we now have a new vector $\vc[\vc{w}]{f}\in\R{\ncol}$ with
\[
\vc[ \vc{w} ]{f}(\vc{x})=\sum_j \vc[][j]{w}\vc[][j]{\breve{x}} \vc
[j]{v}=\mat{V}\mat{D}_{\vc{w}}\vc{\breve{x}},
\]
where $\mat{D}_{\vc{w}}$ is the diagonal matrix with diagonal given by
$\vc{w}$.
For example, if our outside structure could be represented in a
smaller subspace so that~$\bolds{\Sigma}$ had rank \mbox{$r<\ncol$}, then
defining $\vc[][j]{w}=\ind{j\leq r}$ would give $\vc[ \vc{w} ]{f}(\vc
{x})$ as the projection of $\vc{x}$ onto the smaller subspace defined
as relevant by our outside structure. More generally, the eigenvalues
$\lambda_j$ quantify the contribution of a~direction $\vc[j]{v}$ to our
outside structure $\bolds{\Sigma}$, and, therefore, the eigenvalues, or a
monotone transformation of them, are a smoother way to assign relative
importance to the different basis defined by $\bolds{\Sigma}$.

For two vectors $\vc{x}$ and $\vc{y}$, the standard inner-product
between $\vc[\vc{w}]{f}(\vc{x})$ and~$\vc[\vc{w}]{f}(\vc{y})$ is given by
\[
\langle \mathbf{f}_{\vc{w}}(\vc{x}),\mathbf{f}_{\vc{w}}(\vc{y})\rangle = \langle\vc{x},\vc
{y}\rangle_{\mat{V}\mat{D}_{\vc{w}}^2\mat{V}^T},
\]
that is, the inner-product between $\vc{x}$ and $\vc{y}$ using the
metric $\mat{V}\mat{D}_{\vc{w}}^2\mat{V}^T$.
Then the choice of a metric $\rowmetric=\bolds{\Sigma}$ is equivalent to
the choice of weighting each~$\vc[j]{v}$ by $\lambda_j^{1/2}$ and $\vc
[\vc{w}]{f}(\vc{x})=\rowmetric^{1/2}\vc{x}$.

In this light, we can compare the effect of using $\rowmetric=\bolds{\Sigma}$ versus $\rowmetric=\bolds{\Sigma}^{-1}$. Both obviously have
the same eigenvectors and differ only in the weighting the eigenvectors
($\lambda_j$ versus $1/\lambda_j$). Thus, the choice of $\tcov$ as the
metric for the variables places emphasis on the directions with more
information about the outside structure, while $\tcov^{-1}$ emphases
directions that are most independent of the outside information.
Depending on whether this outside structure is thought to enlighten or
confound the analysis, the different weighting systems are appropriate.

From this harmonic perspective, the behavior of the eigenvectors is
quite revealing as to the intuitive interpretation that can be placed
on the analysis. Such a projection onto a relevant set of basis is, of
course, analogous to harmonic analysis or wavelet analysis for
functional data. PCA could also be described similarly, only with the
$\vc[j]{v}$ dependent on the observed variability of the data. In these
cases, the basis functions can be ordered to hopefully reflect
increasingly less meaningful variations of the data, so that the
important information in the data for the analysis in question is
captured in the first few directions. More generally, eigenvectors of a
covariance matrix describe linear combinations of decreasing variance,
and thus presumably decreasing ability to reveal the structure of interest.

Beyond the ordering of the eigenvectors, a desirable behavior for the
purposes of interpretability is for the bases (eigenvectors) to be
sparse---nonzero in a small portion of the coordinate space (or, more
generally, a clearly interpretable subspace). If so, the resulting
coordinates of the transformed data are easily interpreted as contrasts
or combinations of a small set of variables. This is the appeal of
wavelets or various sparse PCA algorithms. From the point of view of
our outside information in the form of a graph or phylogenetic tree,
this means we want our representation of the outside information (via
$\bolds{\Sigma}$) to result in eigenvectors that are interpretable
decompositions of the external information we have. As we will see,
certain covariance structures for phylogenies and also graphs have such
decompositions, which is one reason that the analysis in a nonstandard
inner-product space can give highly interpretable results.

\subsection{gPCA and analyses of variables as observations}\label
{sec:gPCAandDpcoa}
Another interpretation of $\rowmetric$ slightly different from the
geometric one given above is that it is simply an additional data
matrix---one that defines similarities between the $\ncol$ variables---which
we wish to include into our analysis of the primary data
matrix, $\mat{X}$.

\citet{dpcoa} accomplish this by their method of Double Principal
Coordinates Analysis (DPCoA), which explicitly transforms the
similarities between the variables given by $\rowmetric$ into a set of
standard Euclidean coordinates, $\mat{Z}\in\R{\ncol\times r}$, using
MDS (also known as Principal Coordinates Analysis). This can be viewed
as giving an alternative basis for~$\R{\ncol}$ and $\mat{Z}$ as the new
set of coordinates of the original $\ncol$ variables in which~$\mat{X}$
was measured. Then the next step of DPCoA transforms the data~$\mat{X}$
to this new basis as well, that is, to coordinates $\mat{X}\mat{Z}$.
DPCoA then performs PCA on the transformed $\mat{X}$ (we note that
these steps are exactly the same as the steps of DPCoA, but generalized
here to apply to general data matrices~$\mat{X}$ and not just the
contingency tables originally proposed; see Appendix~\ref
{app:DPCoAproof} for details).

The series of steps that make up DPCoA is exactly equivalent to a
single gPCA of the centered data matrix, $\mat{\tilde{X}}$, with the
choice of metrics given by the triplet $(\mat{\tilde{X}}, \rowmetric,
\colmetric)$, provided that (1) the centered data matrix of~$\mat{X}$
was the result of centering the \textit{columns} (variables) and (2) the
same centering matrix used in centering $\mat{X}$ was also used in the
MDS of $\rowmetric$ to find the matrix~$\mat{Z}$ (Appendix \ref
{app:DPCoAproof}). DPCoA was only proposed for the particular setting
of ecological studies where the data is a contingency table, and, thus,
centering the columns of $\mat{X}$ is actually equivalent to centering
the rows because of the row and column weights that are typically
chosen for the centering (see Section \ref{sec:contigProperties}), so
the requirement is naturally satisfied.

By recasting DPCoA as a gPCA, the technique now has general application
and is clearly extendable, since in many situations heterogenous
information can be similarly introduced into an analysis in this way.

We note that MDS is traditionally described based on an input of
squared dissimilarities or distances between points given by a $\ncol
\times\ncol$ matrix $\dist$; however, any positive definite $\rowmetric
$ that can be written as
\[
\rowmetric=\one[\ncol]\vc{v}^T+\vc{v}\mathbf{1}_{\ncol}^T-\tfrac12\dist
\]
for some vector $\vc{v}\in\R{p}$ will result in the same MDS of the
variables and thus the same DPCoA results.

Another approach to analyzing two sources of data are multivariate
kernel techniques, such as kernel CCA [\citet{Bach2002}], which assume
that the only knowledge of the data is similarities between objects. In
these techniques, two sets of data provide two different sets of kernel
similarity matrices $\kernel[1]$ and $\kernel[2]$ on the same set of
$\nrow$ objects, and the kernel analysis results in new coordinates $\vc
[1]{\hat{y}}$ and $\vc[2]{\hat{y}}$ that are linear combinations of
these kernel similarities that best relate the two data sets (the
prediction context is also possible).
Then gPCA of the rows of $\mat{X}$ results in equivalent coordinates
for the rows as the choice of $\kernel[1]=\mat{X}\rowmetric\mat{X}^T$
and $\kernel[2]=\colmetric,$ for an extreme form of regularization of
the CCA problem that only constrains the norm~$\| f \| ^2$ of the
resulting functional, rather than the more common constraint on
estimated variance (see Appendix \ref{app:kernel}).

In the current setting, we are instead interested in outside
information on the $\ncol$ variables in the form of $\rowmetric$. In
this case, the natural kernel analysis would provide new coordinates
for the $\ncol$ columns based on
$\kernel[1]=\mat{X}^T\colmetric\mat{X}$ and $\kernel[2]=\rowmetric,$
which would correspond to a gPCA of the columns. As we noted above,
however, the row coordinates from a gPCA of the rows are recoverable
from the gPCA of the columns. Like DPCoA, this perspective of gPCA is
that of finding a new set of coordinates for the variables, based this
time on explicitly relating the expected similarities to the observed
similarities, and then rotating the matrix $\mat{X}$ into this basis.

\section{Analysis of species abundance}\label{sec:EcoAnalysis}

The investigation of species composition and comparison of species
across different locations, such as in our motivating example of the
bacteria communities, form the core of ecological studies. A large
contingency table of species abundances for different locations is a
common form of data in this literature. Development of gPCA as
described here has often been in this setting, thus it is useful to
review some important points before returning to our bacteria example.

Our motivating example of the bacteria is ecological, but large
contingency tables appear in many other situations. For example, in
document classification, the data could consist of the frequency of
different words in different documents. Another example is allele
frequency studies with the frequency of different alleles of a gene in
different populations. We will continue to focus our notation and
discussion on the phylogenetic/ecological scenario, but the methods
presented here could be of use for these different data types.

\subsection{Notation}\label{sec:notation}
Assume that the abundance of certain species are measured at $\nsites$
different locations and a total of $\nspecies$ distinct species types
are observed. We drop the use of $\nrow$ and $\ncol$ for the rows and
columns of our data matrix to emphasize that there is not a canonical
dimension that is considered the observations in this setting, though
we will focus on the locations as observations in our example. We will
similarly use matrices $\colmetric[\nspecies]$ and $\rowmetric[\nsites
]$ for the row and column metrics.

Let $\ab$ be the resulting $\nsites\times\nspecies$ contingency table
of the observed abundances of species $s$ at location $\ell$. Because
we are interested in comparing the species composition of the
locations, we will represent each location by the relative proportion
of the species in the location. A vector $\locprofile{\ell}$ of
relative proportions at location $\ell$ is called a \textit{profile}
vector in the ecological literature and is obtained by dividing each
row of $\ab$ by its row sum. The corresponding data matrix is given by
$\locmat\in\R{\nsites\times\nspecies}$. Namely, let $\locweightm=\ab\one
/\atotal\in\R{\nsites}$ be the row sums of $\ab$ normalized to sum to
one. Then $\locmat$ is given by
\[
\locmat=
\pmatrix{
\mathbf{x}_{1}^T \cr
\vdots\cr
\mathbf{x}_{\nsites}^T
}
=\wloc^{-1}\ab/\atotal\in\R{\nsites\times\nspecies},
\]
 where $\wloc$ is a diagonal matrix with diagonal elements
given by $\locweightm$ respectively.

The vector $\locweightm$ also defines weights for each of the locations,
and the weights are proportional to the total number of observations in
that location. The weighted mean of the locations, $\locbar$, is given
by $\locmat^T\locweightm$ and the centered data matrix, $\tildlocmat$,
is given by $\tildlocmat=(\mat{I}-\one\locweightm^T)\locmat.$

\subsection{A few important properties of contingency tables}\label
{sec:contigProperties}

\paragraphm{The duality of rows and columns}
Note that the weighted mean, $\locbar$, also sums to one and therefore
is itself a potential location profile. In fact, $\locbar$ is
proportional to the column sums of $\ab$ and thus is equal to the
relative frequency of the species across \textit{all} locations. If we
had instead chosen to analyze the columns (species) as the
observations, choosing weights $\spweightm$ for the species in the same
way as the rows, we would have $\spweightm=\locbar$.

The equivalence of $\spweightm$ and $\locbar$ has interesting
repercussions for data analysis because under these weighting schemes,
we can equivalently center either the rows or the columns,
\[
\tildlocmat=\proj[\locweightm]\locmat=\locmat\proj[\spweightm],
\]
where $\proj[{\vc[m]{w}}]=(\mat{I}_{m}-\one[m]\mathbf{w}_m^T)$ is the
projection matrix that centers $m$ observations based on a weighted
mean with $\vc[m]{w}$ as weights.

\paragraphm{Interpretation of variables in gPCA}
Because we analyze location profiles, there is a simple way to plot the
variables (species) jointly with the observations (locations). Let $\vc
[s]{e}$ be the standard basis vectors of $\R{\nspecies}$. Then $\vc
[s]{e}$ is also a profile vector representing a theoretical location
that consists solely of species $s$. If we transform the data with an
ordination technique, we can jointly transform $\vc[s]{e}$ and plot its
transformation alongside the observed locations. Unlike the usual plots
of variables, the coordinates of our rotated axes have a meaning as a
data point, not just as a direction in space, so we can legitimately
visualize distances between the location and species in a single plot.

\paragraphm{Examples of gPCA with contingency tables}
In addition to DPCoA described above, different metric spaces are often
used for analyzing contingency tables via gPCA, particularly to retain
additional information such as the weights $\locweightm$ and/or
$\spweightm$. The most common example of gPCA is Correspondence Analysis
(CA), which is a gPCA of the row profiles of a contingency table, and
uses the triplet $(\mtilde{\locmat},\wsp^{-1},\wloc)$ [see \citet
{greenacreBook} for a detailed treatment].
This gives an inner product of the form $\mathbf{x}_{k}^T\wsp
^{-1}\locprofile{\ell}$, down-weighting the more frequent species. This
can be seen as counteracting a ``size effect'' for frequencies, where
abundant species dominate the analysis; without this correction,
differences in rare species (which will be on a~smaller order of
magnitude) are lost.

One can argue that the weighting of CA places too much importance on
low abundance species, even though those species are more likely to be
miscounted and are probably less trustworthy. \citet{gimaret1998}
propose no weighting of the species, only the locations, which gives a
triplet $(\tildlocmat,\id[\nspecies],\wloc)$---just a regular PCA with
weights on each observation. Such an analysis in ecology is also called
Non-symmetric Correspondence Analysis (NSCA).

\subsection{Connection to diversity}\label{sec:DivAndGPCA}

We take a moment to comment on the connection of the choice of gPCA
metrics to a common question in ecology---how ``diverse'' a location
is. Diversity is a measurement of how close the distribution of species
is to uniform. Two popular measures of diversity are variations of the
Gini--Simpson index, $\divgs(\locprofile{})=1-\sum_{s=1}^\nspecies
\locprofile[s]{}^2,$ and the Shannon Diversity index, $\divsh
(\locprofile{})=\sum_{s=1}^\nspecies\locprofile[s]{}\log(\locprofile[s]{})$.

Ecology studies often use the individual diversity of locations to make
comparisons, but the diversity indices alone do not effectively compare
the species composition. Locations can have quite different composition
of species but with same levels of individual diversity. Of interest is
how the species composition changes, and ordination techniques are used
to address these problems, but as a separate component of the analysis
of the ecological data. However, the choice of diversity and the choice
of gPCA parameters are closely connected, as pointed out in \citet
{Pelissier}. Namely, if $\colmetric[\nsites]$ is a simple diagonal
matrix of weights on the locations, gPCA of $(\mat{X}, \rowmetric
[\nspecies],\colmetric[\nsites])$ gives the best representation of a
particular measure of dissimilarity between locations, and choice of
this dissimilarity measure implies a diversity measure, and vice versa.
\citet{Pelissier} stated this for several specific ordination
techniques, and we state it more generally for any choice of metric
$\rowmetric[\nspecies]$ on $\R{\nspecies}$. Define diversity and
dissimilarity measures for any positive definite matrix $\rowmetric
[\nspecies]=\rowmetric[]$ by
\begin{eqnarray*}
H_{\rowmetric[]} (\locprofile{})
&=& \locprofile{}^T \operatorname{diag}(\rowmetric[]) -
\locprofile{}^T \rowmetric[] \locprofile{}=
\sum_r \locprofile[r]{}\mat[(rr)]{\rowmetricLetter}-\sum_{rs}\mat
[(rs)]{\rowmetricLetter} \locprofile[r]{}\locprofile[s]{},
\\
\operatorname{Diss}_{\rowmetric[]}(\locprofile{k},\locprofile{j})&=&
(\locprofile{k}-\locprofile{\ell})^T\rowmetric[](\locprofile
{k}-\locprofile{\ell}).
\end{eqnarray*}
These are clearly closely related to the norm and inner-product defined
with the choice of $\rowmetric[]$. With these choices of diversity and
dissimilarity,\vadjust{\goodbreak} the total diversity across all locations is given by
$H_{\rowmetric[]}(\locbar)$ and can be decomposed into the average
diversity of individual locations and plus the average of pairwise
dissimilarities of locations,
\[
\underbrace{\vphantom{\sum_{\ell=1}^L}H_{\rowmetric[]}(\locbar
)}_{I_{\mathrm{Total}}}
= 1/2 \underbrace{\sum_{k=1}^L \sum_{\ell=1}^L
\locweight[k]\locweight[\ell] \operatorname{Diss}_{\rowmetric[]}(\locprofile{\ell
},\locprofile{\ell})}_{I_{\mathrm{Between}}}+\underbrace{\sum_{\ell=1}^L
\locweight[\ell] H_{\rowmetric[]}(\locprofile{\ell})}_{I_{\mathrm{Within}}} .
\]
gPCA of $(\mat{\tilde{X}}, \rowmetric[],\mat{D}_{\locweightm})$ gives
the best low-dimensional representation of $I_{\mathrm{B}}$, the average
dissimilarity between locations (see Appendix \ref{app:decomposeDiv}).

We can define a $F$-style statistic, as in ANOVA, to test for
significant dissimilarity between the locations [\citet{LegendreBook}]
\[
F=\frac{(\atotal-1) I_{\mathrm{B}} }{L I_{\mathrm{W}}}.
\]
Because the significance of $F$ will generally be determined by
permutation tests, this $F$-test is functionally equivalent to using
$I_{\mathrm{B}}/I_{\mathrm{T}}$, which has many appealing connections to standard measures.
We describe a few of them below given originally by \citet{Pelissier}
and \citet{dpcoa}:
\begin{description}
\item[CA:]For correspondence analysis, $\rowmetric[]=\wsp^{-1}$
results in a dissimilarity between profiles measured by the $\chi^2$ distance,
\[
(\locprofile{k}-\locprofile{\ell})^T\wsp^{-1}(\locprofile{k}-\locprofile
{\ell}),
\]
which has also been proposed for document classification. As is well
known in CA, $I_{\mathrm{B}}=\chi^2/\atotal$, where $\chi^2$ is the $\chi
^2$-statistic for testing independence. The implied diversity
measurement for a profile $\vc{x}$ is \mbox{$\sum\spweight[r]\locprofile
[r]{}(1-\locprofile[r]{})$}, which implies the total diversity $I_{\mathrm{T}}$ is
simply $\nspecies-1$. Thus, $I_{\mathrm{B}}/I_{\mathrm{T}}$ is proportional to the $\chi^2$ statistic.
\item[DPCoA:] As we saw before, DPCoA can be written in terms of a
general $\rowmetric[\nspecies]$. If we write $\rowmetric[\nspecies]=\one
[\ncol]\vc{v}^T+\vc{v}\mathbf{1}_{\ncol}^T-\half\dist$ for some $\vc{v}\in\R
{\nspecies}$ and species dissimilarities~$\dist$, as in Section \ref
{sec:gPCAandDpcoa}, then we have that $H_{\rowmetric[]}$ and
$\operatorname{Diss}_{\rowmetric[]}$ are the
Rao diversity and dissimilarity measures [\citet{RaoDiv1982}] given by
\begin{eqnarray*}
H_{\rowmetric[]} (\locprofile{})&=&\sum_{rs}\dist_{(rs)} \locprofile
[r]{}\locprofile[s]{},\\
\operatorname{Diss}_{\rowmetric[]}(\locprofile{k},\locprofile{j})&=&
(\locprofile{k}-\locprofile{\ell})^T\bigl(-\tfrac{1}{2}\dist\bigr)(\locprofile
{k}-\locprofile{\ell}).
\end{eqnarray*}
Thus, gPCA with $\rowmetric[\nspecies]$ results in differences between
locations profiles being down-weighted for the species that are similar
to each other and up-weighted for very distinct species. Though stated
in many individual steps and not a single gPCA as we do here, the DPCoA
method was motivated by searching for an ordination that maximized this
notion of distance between observations. The ratio $I_{\mathrm{B}}/I_{\mathrm{T}}$ is
commonly called the $F_{\mathrm{ST}}$ statistic [\citet{MartinA2002}] in
biological applications and has been suggested for testing differences
in bacterial communities, where $\dist$ is usually chosen as the
original measures of genetic distance between the sequences. The
$F_{\mathrm{ST}}$ statistic is also used in testing for differences of allele
composition in human populations [\citet{Excoffier1992}].
\item[NSCA:] Since NSCA is standard PCA, except for the weighting of
the observations, $\rowmetric[]=\id,$ and is equivalent to the Rao
diversity and dissimilarity measures when all the species are equally
distant from each other. The resulting measure of diversity in this
case is the Gini--Simpson measure of diversity, $\divgs$. The ratio
$I_{\mathrm{B}}/I_{\mathrm{T}}$ is equivalent to Kendall's $\tau$ [\citet{DAmbra1992}].
\end{description}

\section{A metric for species related by a phylogenetic tree}\label
{sec:phylometric}
Returning to our bacteria example, we want a matrix $\bolds{\Sigma}$ that
represents the phylogenetic relationships of the species. As mentioned
in Section \ref{sec:generalApproach}, if we can model the covariance
structure of data expected based on just our outside information, this
provides a natural choice of $\bolds{\Sigma}$. The phylogenetic tree in
fact is a representation of the process of evolution, for which many
possible probabilistic models could be created.

A common probabilistic model for the evolution of the value of a trait
over time, due to \citet{cavalli1975}, is one of a Brownian motion
model over time, where at each speciation event the model assumes that
the resulting sister species continue to evolve independently [for
alternative models of evolution, see \citet{Hansen1996}; \citet{Pavoine2008}].
This model gives a covariance structure for the trait as observed on
the existing species (the leaves of the phylogenetic tree) and can be
simply stated in terms of distances between species on the phylogenetic
tree. Moreover, the eigenvectors of this covariance matrix generally
demonstrate nice localization properties relative to the tree, implying
interpretable results in terms of the properties of the tree.

Specifically, assume that there is a known phylogenetic tree describing
the ancestral relationship of $\nspecies$ extant species and that a
trait of interest for these species has evolved over time according to
the model of independent Brownian motion with the speciation as
depicted on this tree. The $\nspecies$ extant species are observed, and
for each species $s$ at a single time point $\vc[][s]{t}$, the trait is
measured, resulting in $\vc[][s]{y}$. Then the vector of trait values,
$\vc{y}$, follows a multivariate normal distribution with covariance
between species $r$ and $s$ proportional to the total length of time
that the evolutionary history of the two species were identical,
$\cov(\vc[][s]{y},\vc[][r]{y})= \sigma^2 t_{rs},$
where $t_{rs}$ is the time at which the two lineages diverged, as
measured from their most common ancestor.

We can write this covariance quite simply in terms of the topology of
the tree and the length of the branches, assuming that the branch
length is reflective of evolutionary time. Let $\dist$ be the distance
matrix of the leaves based on the distance of the shortest path between
them on the tree. Then we can write the covariance matrix $\tcov$ as
\[
\tcov=1/2(\one\myvec{t}^T+\myvec{t}\one^T-\dist),
\]
where $\vc{t}\in\R{\nspecies}$ is the vector of the distance of each
species to the root.

This relationship between $\tcov$ and $\dist$ implies that gPCA with
$\tcov$ as the species metric will decompose a Rao Dissimilarity, with
dissimilarities between species given as their distance on the tree.
For the bacterial example, use of this distance has the effect of not
declaring locations very different if the differences between locations
occur in phylogenetically similar phylotypes.

\paragraphm{Properties of phylogenetic metric}\label{sec:phyloproperties}
We would like that the eigenvectors of~$\tcov$ be sparse in a useful
way relative to the structure of the tree, for example, that they
contrast sister subtrees of the phylogenetic tree and be zero
elsewhere. Furthermore, we would like that eigenvectors give
increasingly specific level of detail so that eigenvectors
corresponding to larger eigenvalues highlight deeper structure in the
tree. Put together, these statements would imply that the eigenvectors
offer a multiscale analysis of the tree, with eigenvectors
corresponding to large eigenvalues interpretable as summarizing
differences in the large initial partitions of the tree and smaller
eigenvalues giving eigenvectors reflecting the distinctions between the
later divisions of the tree.

\begin{figure}

\includegraphics{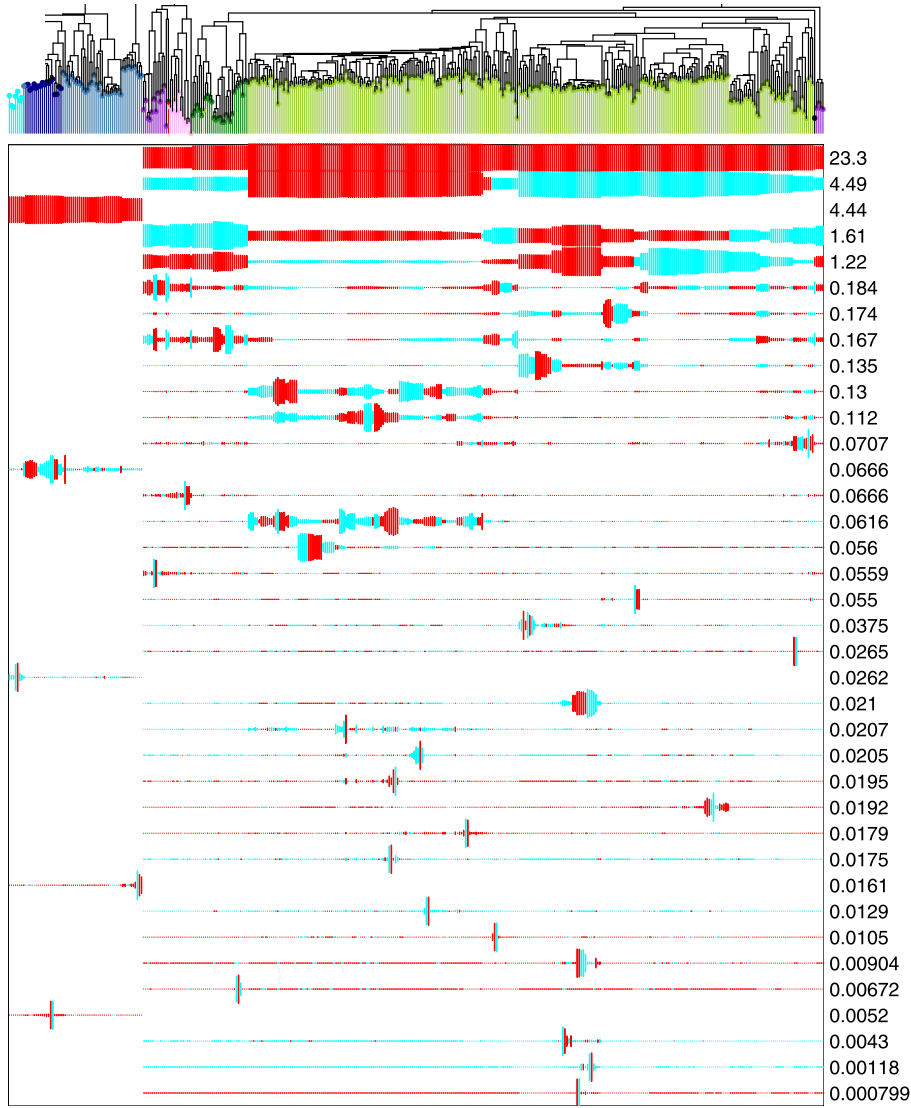}

\caption[Eigenvectors and eigenvalues of $\tcov$ for intestinal
data]{An illustration of some eigenvectors of $\tcov$ for the
intestinal data. Only 25 of the 395 eigenvectors are shown: those that
correspond to the first five largest eigenvalues, the last two smallest
eigenvalues, and then a random sample in between. Each row represents
an eigenvector, and the value of each element of the vector is plotted
alongside the phylotype with which it corresponds. Blue represents a
positive value, red a negative. The width indicates the absolute value
of the element. Again, each row has been normalized so that the maximum
width is the same in each row. Next to each row is printed the
corresponding eigenvalue.}
\label{fig:intesEV}
\end{figure}

Several authors in phylogenetics have asserted that the eigenvectors of
$\tcov$ have this multiscale structure [e.g., \citet{cavalli1975}; \citet{rohlf2001}; \citet{martins2002}], but only limited statements of this kind can
be rigorously made about a phylogenetic tree with more than four
leaves/species [see \citet{purdom2006} for a longer discussion]. But
empirical observations of the eigenvectors show that they often do have
some characteristics of this multiscale property; for example, $\tcov$
has a block structure which guarantees that the eigenvectors of $\tcov$
will, at a minimum, be nonzero for only one side or the other of the
initial split in the tree (Appendix \ref{app:eigenvectors}).
Beyond this, if we ignore the comparatively small values in the
eigenvector, eigenvectors corresponding to smaller eigenvalues do tend
to divide the species into smaller and smaller closely-related groups
based on the sign of the entries, though the groups do not exactly
correspond to subtrees (see Figure \ref{fig:intesEV}).

\section{gPCA applied to bacterial data and phylogenetic tree}

\begin{figure}
\centering
\begin{tabular}{@{}c@{\hspace*{3pt}}c@{}}

\includegraphics{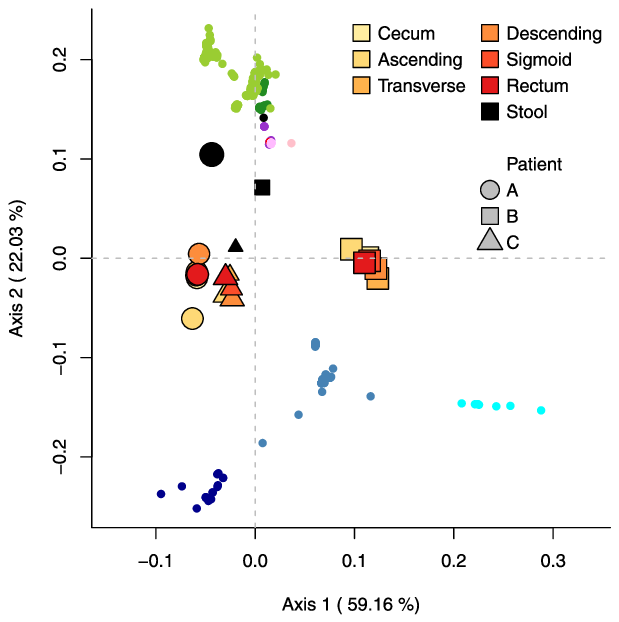}
&\includegraphics{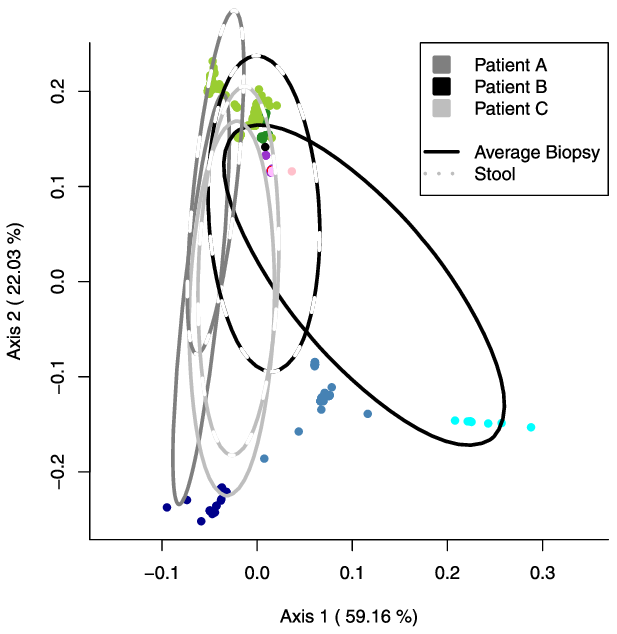}\\
\footnotesize{(a)}&\footnotesize{(b)}
\end{tabular}
\caption{Scatter plot of the
species and samples with the first two coordinates given by DPCoA.
Species are shown as colored points in both plots. In plot \textup{(a)}, the samples are shown as the large blue shapes:
different shapes indicate different patients and different shades of
blue indicate location within the colon. In plot \textup{(b)}, samples are represented as ellipses that
indicate the major directions of the abundances of the samples. For
simplicity, a single ellipse for the combined abundance in the biopsies
is shown because the internal biopsies are very similar.}
\label{fig:intesDPCoA}
\end{figure}

In \citet{myscience} our original analysis of the bacterial data was a
gPCA of $(\mat{\tilde{X}}, \bolds{\Sigma},\mat{D}_{\locweightm})$, which
is equivalent to DPCoA choosing $\dist$ to be the distance among the
phylotypes. We display in Figure \ref{fig:intesDPCoA} the ordination of
the locations (samples) and species using the first two coordinates
(using the implementation of DPCoA in the \texttt{ade4} package in R
[\citet{Rade4}; R~Development Core Team (\citeyear{R})]). The first obvious fact is that the patients are
separated, almost entirely, by just their value when projected onto the
first axis. The first axis orders the patients B, C, A, which correlates
with visual examination of the data in Figure \ref{fig:intesData}.
Below we will compare to other common choices of metrics and we will
see that distinguishing the patients is not difficult since all of the
techniques accomplish this, though not always in just one dimension.
More interestingly, we also see in Figure \ref{fig:intesDPCoA} that the
stool samples are distinguished from the internal biopsies of the
colon, and the second axis seems to make this distinction. Again this
makes sense from visually examining the data, since within each patient
the stool samples do stand out from the biopsies.

The most striking aspect of the plot from the gPCA is the additional
information provided from the inclusion of the phylotypes in the plot.
Recall that when our data matrix $\mat{X}$ consists of profile vectors,
our original axes $\vc[s]{e}$ correspond to a location that is entirely
concentrated in phylotype $s$. The coordinates of the phylotypes given
by gPCA will be the coordinates of our axis $\vc[s]{e}$ centered and
rotated like the observed profiles (see Appendix~\ref{app:DPCoAproof}).
Looking at the ordination plot, we see that the phylotypes' coordinates
provide an interpretation for the first two dimensions. The phylotypes
are in clusters much like the groupings on the tree---not surprising
if we recall that in the full space the distances between the species
are exactly the distances on the tree. What is interesting is how the
clusters on the tree fill the space once projected into these two
coordinates that preserve the Rao Dissimilarity among the locations.
The distribution of the phylotypes indicate the importance of these
clusters in determining the dissimilarity between the patients. Those
far from the origin have more impact in defining the coordinates of the
locations. We see the tension between the various \textit{Bacteroides}
(blue) and the rest of the tree.

Furthermore, we can interpret the relationship between the locations
and the phylotypes. We see that patient B is comparatively much more in
the direction of the \textit{Prevotallae}-like bacteria (light blue),
while the other two patients are more in the direction of the \textit{B.
Vulgatus}-like phylotypes (dark blue). Similarly, the biopsies are
comparatively more heavily represented in the \textit{Bacteroides} (blue)
portion of the tree, while the stool samples are comparatively less so.
Figure \ref{fig:intesDPCoA}(b) depicts the different samples as
ellipses with the axes of the ellipses determined by the relative
proportion of the different species for the location (see Appendix \ref
{app:ellipses}). This illustration emphasizes that the samples can be
thought of giving weights to each phylotype, and the ellipse
demonstrates the relative influence of the different species. We see
graphically the different influences of the two groups of \textit
{Bacteriodes} (blue) in separating the biopsies of patient B from all
of the rest of the samples. Transforming the data in various ways
before analysis does not dramatically change these relationships (e.g.,
log-transforming the data or adding pseudo-counts).

\begin{figure}
\centering
\begin{tabular}{@{}c@{\hspace*{7pt}}c@{}}

\includegraphics{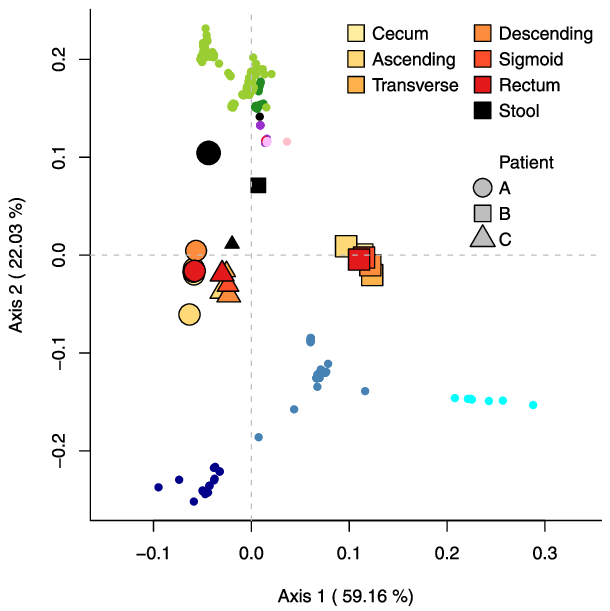}
&{\includegraphics{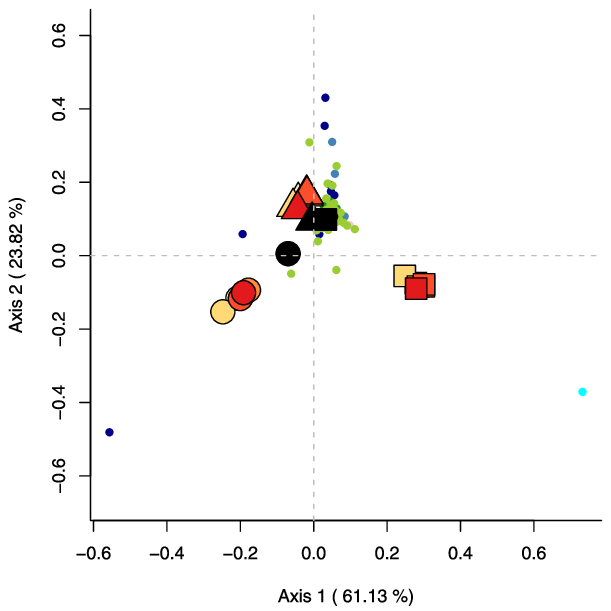}}\\
\footnotesize{(a) gPCA with Tree / DPCoA}&\footnotesize{(b) NSCA (Gini--Simpson Distance)}\\[6pt]

\includegraphics{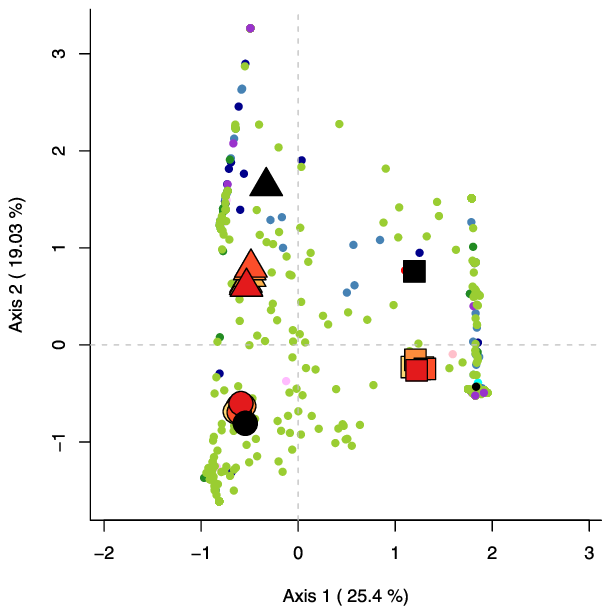}
&{\includegraphics{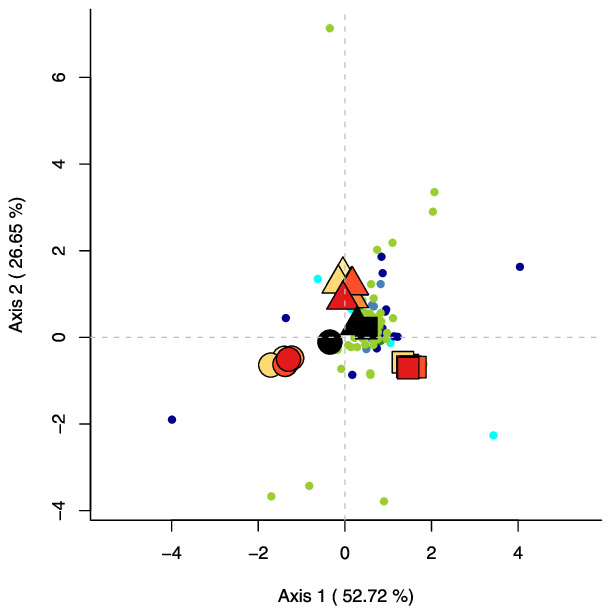}}\\
\footnotesize{(c) CA}&\footnotesize{(d) gPCA with $\tcov^{-1}$}
\end{tabular}
\caption[Other ordination techniques on the bacterial data]{Coordinates
of species and samples from alternative ordination techniques.}
\label{fig:intesOtherMethods}
\end{figure}

All of these visualizations have, by necessity, focused on only the
first two dimensions of the coordinates given by gPCA. These dimensions
do cover a~large proportion of the Rao Dissimilarity, but still are
only an approximation of the full space. We are mainly focused on
demonstrating the characteristics of the ordination procedure in terms
of the coordinate system that it creates, but for more rigorous testing
of differences between the patients or between the biopsies and stool
samples, permutation tests based on the $F$-statistic described above
would generally want to compare with the entire coordinate system.

\subsection{Comparison to other approaches}
How do these results compare to the other ordination techniques
mentioned above? In Figure \ref{fig:intesOtherMethods} we show the
results of the ordination from Non-symmetric Correspondence Analysis
(NSCA), Correspondence Analysis (CA) and a Mahalanobis-like distance
based on $\tcov^{-1}$ (see Section~\ref{sec:phylometric}). We similarly
center, rotate and project the axes $\vc[s]{e}$ to get the species
coordinates in the same manner as DPCoA.

As we mentioned, all of the techniques separate the three patients, but
we see that the gPCA using the tree gives much more relevant results,
both in terms of the role of the species and in relating to our
intuitive interpretation of the data. The NSCA [plot (b)] is the same technique as our gPCA but with
each species at equal distance from every other; it is also just a
standard PCA with weights on the observations. In the first two
coordinates of the NSCA, we see that instead of having a smooth
contribution from clusters of phylotypes, two individual phylotypes,
far removed from the rest, contribute to the division of the patients
much more than the rest. The bulk of the species have little
contribution to these coordinates. Thus, there is little from which to
draw more general conclusions regarding the biological characteristics
of the species which are influential. This is a consequence of treating
each phylotype equally, rather than using the additional structure of
the tree to shape the analysis. CA [Figure \ref{fig:intesOtherMethods}(c)],
on the other hand, spreads out the importance of each phylotype. Here
we can see the effect of the down-weighting metric in CA discussed
earlier; differences found in the many low abundance phylotypes are
allowed to influence the analysis. Rather than a couple of phylotypes
dominating the analysis, as in NSCA, the phylotypes play more equal roles.

We might try to use any one of these techniques to reason out
relationships among the variables. Each technique would give a
different story in the role of the variables (phylotypes) dependent
upon the assumptions inherent in the method. The relevant feature for
our analysis is that we presuppose that a certain type of information
is relevant---namely, how the structure of the tree relates to the
data. This approach focuses the analysis on finding an interpretation
among the variables that follows the tree structure.

We note that the abundance table from metagenomic studies discussed
here has many features common to high-throughput experiments in biology---in
particular, the number of biological samples is quite low
compared to the number of measurements. We sought to integrate the
phylogenetic information into the data analysis \textit{a priori}. In
this way, the analysis is constrained in a biologically relevant
direction. In contrast, we could think of analyzing this abundance data
much like a microarray experiment: test each phylotype individually for
differences between the patients and use multiple testing criteria to
identify individual phylotypes showing significant differences. A~%
problem with this approach, which is also a common problem in
microarrays analyses, would be that a list of significant phylotypes is
difficult to interpret. In microarray studies, biological
interpretation is often done \textit{a~posteriori} by then examining
biological knowledge of the list of genes. We could similarly use the
phylogenetic tree in this way. However, we just saw that an analysis
independent of the tree highlighted only a couple of specific
phylotypes from which it would be difficult to build a general
connection to the tree.

\subsection{Effect of the choice of metric}

We can see the effect of using $\tcov$ in our gPCA by examining the
linear combinations that gPCA using $\tcov$ chooses. For any ordination
technique, let $\mat{V}$ be a matrix that rotates the \textit{original}
profiles~$\locmat$ to give us the final ordination; in gPCA of centered
data, this will be the matrix $\proj[\spweightm]\rowmetric[S]\mat{A}$.
We examine the different linear transformations, $\vc[i]{v}$, from gPCA
with $\tcov$ as compared to the transformation for a standard PCA on
the data $\tildlocmat$ (equivalently, NSCA). And we also compare to the
eigenvectors of $\tcov$: if the covariance between the species was
exactly the $\tcov$ predicted by the evolution model, then these would
be the principal components of such data. Thus, we can think of the
eigenvectors of $\tcov$ as PCA on the tree.

\begin{figure}

\includegraphics{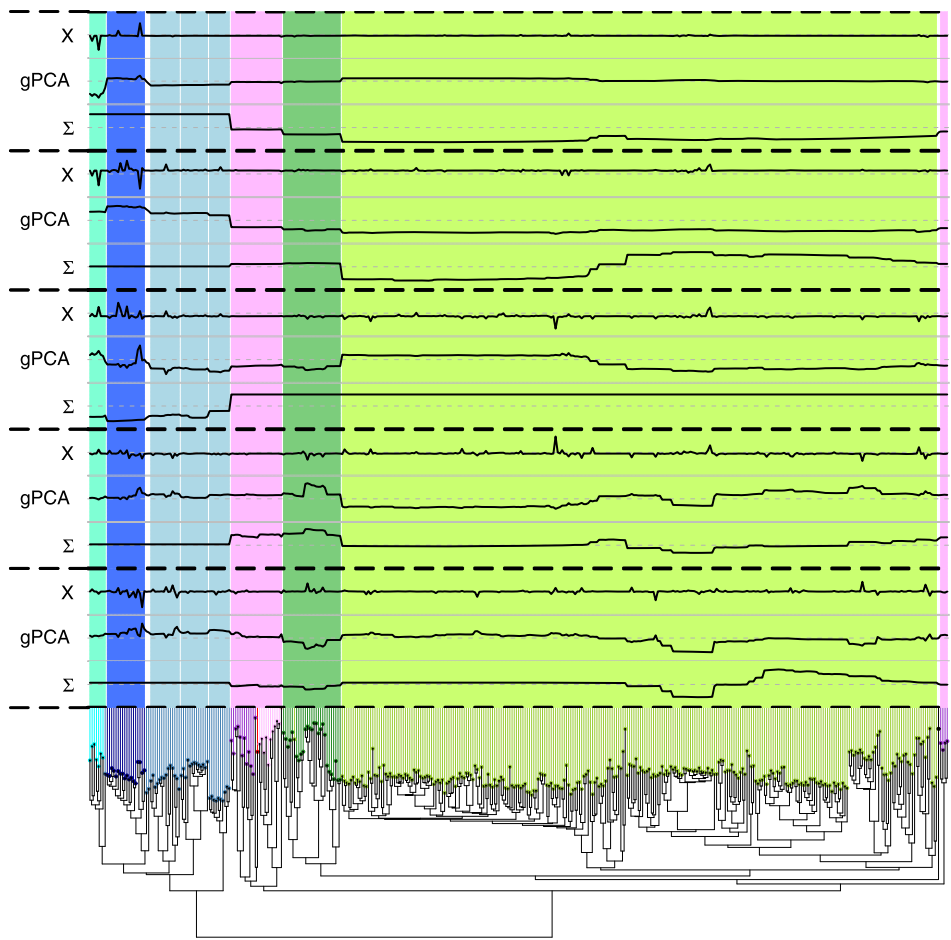}

\caption{Shown are the first five linear
combinations of gPCA using $\tcov$ that act on the observations in
$\locmat$ (the location profiles) to create the first five coordinates
($\vc[i]{v}$). The five dimensions are divided by thick, dotted line.
Also shown adjacent to each gPCA vector are the linear combinations
from a standard PCA of $\tildlocmat$ (labeled `X') and the eigenvectors
of~$\tcov$ (labeled `$\tcov$').}
\label{fig:intesEVSmooth}
\end{figure}

In Figure \ref{fig:intesEVSmooth} we order the elements of $\vc[i]{v}$
from these three ordination techniques so that they line up with the
phylogenetic tree. In this way we can see the relative importance of
the phylotypes in transforming the data. When we look at the linear
combinations for the first few coordinates, we see that the principal
components from our gPCA with $\tcov$ intuitively seem to be a~%
trade-off between these two options, and we could think of this as a
shrinking of the data variability in the ``direction'' of the tree. This
is a particularly appealing idea, since we are treating the phylotypes
as variables and there are far too many variables for the number of
samples we have.

Despite the intuitive results, the analysis depends on our choice of
encoding the tree using $\tcov$ (or, equivalently, for DPCoA, our
choice of $\dist$). In particular, the block structure of $\tcov$ puts
large emphasis on the first initial partition of the species at the
root of the tree; these two groups of species are considered
independent, conditional on the root ancestor. We can see this emphasis
on this first divide from the Rao Dissimilarity based on $\dist$, where
these two lineages will be far away from each other and, thus,
differences between will be accorded more weight in the analysis.
However, as mentioned above, we see that the method depends only on~$\dist$, so the definition of the root of the tree, per se, is not the
deciding factor, but rather the large amount of distance between these
two subtrees.

Changes near the tips of the tree, both in the numerical data and the
definition of the tree, will have little impact on the gPCA. For the
bacterial data that we are interested in, the deeper tree structure is
more trustworthy than the structure near the leaves of the tree because
of the approximate definition of species. It is a reasonable compromise
to put more weight on the deeper structure of the tree, and base our
analysis on this dependence, in exchange for resolving the more
fundamental problem in our definition of the species.

\section{General graphs}\label{sec:generalGraphs}

It is clear that the same approach is applicable to other situations
where there is complicated information that is related to the
experimental data. By understanding our phylogenetic analysis as a
specific example in a general approach to data analysis, we can compare
with other techniques as well as take advantage of insights from other
data situations.

A closely related example is when we have not a phylogenetic tree, but
a~more general graph structure that describes the relationship of our
variables or observations. The analysis of experimental data in tandem
with related biological networks by \citet{Rapaport2007} is equivalent
to our metric approach. There, the authors used the Laplacian matrix
associated with a~graph to represent the biological graphs that related
genes, where the the laplacian matrix $\mat{L}$ is given by $\mat
{D}_{\vc{d}}-\mat{A}$, where $\mat{A}$ is the adjacency matrix of the
graph and $\vc{d}$ is the vector of degrees of each node. The Laplacian
matrix is a natural choice for graphs; the eigenvectors have similar
multiscale properties as our metric for the phylogenetic tree. In
Appendix \ref{app:treeLaplacian} we briefly discuss the possibility of
treating the phylogenetic tree as a general graph and using the
Laplacian as a metric. We chose another approach here because such a
choice does not well reflect the phylogenetic information in the
tree.\vadjust{\goodbreak}

A related application is found in spatial analysis, where spatial
relationships between observations are based on neighborhood
relationships, or, more generally, distances between points. While many
analyses first remove the spatial dependencies so as to have
independent observations, it is often also of interest to evaluate the
relationship between the spatial patterns and the observed data. When
spatial connectivity between observations is simplified to a zero--one
connectivity measure (usually based on a cutoff on the distance between
the observations), the spatial relationship is given by an adjacency
matrix. Geary's $c$ and Moran's $I$, two common measures of the spatial
autocorrelation of $\vc{y}\in\R{n}$ (a variable observed on the $n$
observations), can be written in terms of the adjacency matrix [\citet
{Thioulouse1995}],
\begin{eqnarray*}
c&=&\frac{(n-1)\sum_{j=1}^n\sum_{j=1}^n \mat[(ij)]{A}(\vc[][i]{y}-\vc
[][j]{y})^2}{2 N_e \sum_i (\vc[][i]{y}-\bar{y})^2}
=\frac{n-1}{N_e}\frac{\vc{\tilde{y}}^T(\mat{D}_{\vc{d}}-\mat{A})\vc
{\tilde{y}}}{\vc{\tilde{y}}^T\vc{\tilde{y}}},\\
I&=&\frac{(n)\sum_{j=1}^n\sum_{j=1}^n \mat[(ij)]{A}(\vc[][i]{y}-\bar
{y})(\vc[][j]{y}-\bar{y})}{2 N_e \sum_i (\vc[][i]{y}-\bar{y})^2}=\frac
{n}{N_e}\frac{\vc{\tilde{y}}^T\mat{A}\vc{\tilde{y}}}{\vc{\tilde{y}}^T\vc
{\tilde{y}}},
\end{eqnarray*}
where $\vc{\tilde{y}}=\vc{y}-\bar{y}\one[n]$ is $\vc{y}$ centered by
the standard (unweighted) mean of the elements of $\vc{y}$ and $N_e=\sum
_{ij}\mat[ij]{A}$ is twice the number of edges in the graph. In
particular, we see that Geary's $c$ can be written in terms of an
inner-product using the Laplacian.

\citet{Thioulouse1995} note that the variance of $\vc{y}$ with
observations weighted by their node-degree, given by $\var_{\mat{D}_{\vc
{d}}}(\vc{y})=\vc{\tilde{y}}^T\mat{D}_{\vc{d}}\vc{\tilde{y}}/N_e$, can
be decomposed into related components,
\[
\var_{\mat{D}_{\vc{d}}}(\vc{y})=\vc{\tilde{y}}^T\frac{\mat{D}_{\vc
{d}}-\mat{A}}{N_e}\vc{\tilde{y}} + \vc{\tilde{y}}^T\frac{\mat
{A}}{N_e}\vc{\tilde{y}}.
\]
Thus, Geary's $c$ and Moran's $I$ are similar to $F$ measures described
above, that is, the ratio of component variability to total variability
(note, however, that Moran's $I$ can be negative). Several authors have
proposed spatial multivariate analyses which rely on $\mat{L}$ as a
metric for the rows, or a row standardized version $\mat{L}^{*}=\mat
{D}_{\vc{d}}^{-1}(\mat{D}_{\vc{d}}-\mat{A})$ [\citet
{AlujaGanet1991p3648}; \citet{Thioulouse1995}; \citet{diBella1996p3649}; \citet{Dray2008}].
[Note that the matrix $\mat{L}^*$ as well as the similar matrix $\mat
{\widetilde{L}}=\mat{D}_{\vc{d}}^{-1/2}(\mat{D}_{\vc{d}}-\mat{A})\mat
{D}_{\vc{d}}^{-1/2}$ are also considered in graph theory; see \citet
{Biyikoglu2007}.]

\section{Conclusion}

There is a clear necessity for including phylogenetic information in an
analysis of metagenomic data. gPCA gives a simple and compelling way to
accomplish this. We also see from our recasting of DPCoA as a gPCA that
the framework of gPCA allows for easy comparisons between seemingly
disparate analyses as well as further exploration as to the effect of
our choice of metrics.\vadjust{\goodbreak}

The use of nonstandard metrics is quite natural in statistics and can
be implemented in a variety of ways, PCA being merely the simplest.
Common examples, such as Mahalanobis distance, are usually data-driven,
but we see that metrics based on outside knowledge can be used to
include complicated and heterogeneous information into the analysis of
our numerical data. This kind of information can help to give more
context to the data, particularly when the number of variables is large
as compared to the samples. Moreover, since the metrics here correspond
to covariance matrices, probabilistic models give a simple approach for
encoding information appropriately. Often, as in the case of
phylogenetic trees, the eigenvectors of such covariance matrices have
nice localization properties that highlight the relevant spatial or
regional patterns of the prior information.

\begin{appendix}

\section{DPCoA and gPCA}\label{app:DPCoAproof}
We state here the equivalence between DPCoA and gPCA described in Section~\ref{sec:gPCAandDpcoa}. First we describe more explicitly DPCoA, as
described in \citet{dpcoa}.

\paragraphm{DPCoA} Assume that the \textit{squared} pairwise
distances/dissimilarities between the species are given by a $\nspecies
\times\nspecies$ matrix $\dist$. We also assume that the distances are
Euclidean (i.e., coordinates can be found for the points so that the
standard Euclidean distance between points is given by the square-root
of the entries of $\dist$).

Following the notation provided in Section \ref{sec:EcoAnalysis}, let
$\proj[{\vc[m]{w}}]=(\mat{I}_{m}-\one[m]\mathbf{w}_{m}^T)$ be the projection
matrix that centers $m$ observations based on a weighted mean with $\vc
[m]{w}$ as weights:
\begin{enumerate}[3.]
\item Find Euclidean coordinates of the species using a weighted
version of Multidiminsional Scaling, with weights for the species given
by $\spweightm$, typically [and as proposed by \citet{dpcoa}] the
relative abundance of the species in all the samples. Specifically, let
$\mat{U}$ be the eigenvectors of
\[
\wsp^{1/2}\proj[\spweightm](-\dist/2)\proj[\spweightm]^T\wsp^{1/2}.
\]
Then the new coordinates of the species are given by the rows of $\dpx
\in\R{\nspecies\times s^*}$ ($s^*\leq\nspecies-1$ is the dimension of
the space required to contain the species). Then we have $\dpx=\wsp
^{-1/2}\mat{U}\bolds{\Lambda}^{1/2}$.
Note that we could also start with a similarity matrix between species,
$\similar=\one\vc{v}^T+\vc{v}\one^T-\half\dist$ for any~$\vc{v}$ that
implies $\similar$ is positive definite. Because
\[
\proj[\vc{w}]\similar\proj[\vc{w}]^T=\proj[\vc{w}](-\dist/2)\proj[\vc{w}]^T
\]
for any weights $\vc{w}$ and vector $\vc{v}$ the MDS will be
equivalent. This is, of course, the standard equivalence between
starting with a similarity matrix or dissimiliarity matrix in MDS.
\item Set the coordinates of the locations to be at the barycenter of
the species coordinates. In other words, each location $\ell$ is given
coordinates that are the weighted average of the coordinates of all the
species and the weights are given by the relative abundance of the
species in that site (which is contained in the vector $\locprofile{\ell
}$). Let the rows of the $\nsites\times s^*$ matrix $\dpy$ contain the
coordinates of the sites, so
\[
\dpy=\locmat\dpx.
\]
The squared pairwise Euclidean distance between the locations using
these coordinates will be equal to their Rao Dissimilarity using the
dissimilarity matrix~$\dist$.
\item Find a lower-dimensional representation of the locations using a
generalized principal components analysis on the triplet $(\dpy,\id
[\nspecies],\wloc)$, where~$\wloc$ is a diagonal matrix consisting of
weights for the locations, $\locweightm$ (again, typically the relative
abundance of the locations in all the samples). Let $r=\operatorname{rank}(\dpy)$.
Then gPCA of $(\dpy,\id[\nspecies],\wloc)$ gives the eigenvalue equations,
%
\begin{eqnarray}\label{eq:DPCoAeigeneq}
\dpy^T\wloc\dpy\mat{F}=\mat{F}\bolds{\Phi}, \qquad
\dpy\dpy^T\wloc\mat{G}=\mat{G}\bolds{\Phi},\nonumber
\\[-8pt]
\\[-8pt]
 \eqntext{\mbox{where }
\mat{F}^T\mat{F}=\mat{I}_r,\
\mat{G}^T\wloc\mat{G}=\mat{I}_r}
\end{eqnarray}
 and $\dpy=\mat{G}\bolds{\Phi}^{1/2}\mat{F}^T$ is the
generalized SVD decomposition of $\dpy$.
The final coordinates of the locations are given by
\[
\sitecoord=\dpy\mat{F}.
\]
We also transform the coordinates of the species to get species
coordinates (see Section \ref{sec:contigProperties}),
\[
\speciescoord=\dpx\mat{F}.
\]
\end{enumerate}

\begin{lem*} \label{thm:DPCoA-gPCA} The coordinates for the locations
given by $\sitecoord$ in DPCoA using\vspace*{1pt} $\dist$ are equivalent to the
coordinates $\mhat{X}=\tildlocmat\similar\mat{A}$ of the locations
given by gPCA with the triplet $(\tildlocmat,\similar,\wloc)$, where
$\tildlocmat=\mat{X}\proj[\spweightm]$ is the \textit{column} centered
matrix of data. Furthermore, the coordinates of the species given by
DPCoA in the matrix $\speciescoord$ are equivalent to the coordinates
obtained by centering and then rotating the original axes $\vc[s]{e}$
by the transformation implied from the gPCA of $(\tildlocmat,\similar
,\wloc)$ so that
$\speciescoord= \proj[\spweightm]\similar\mat{A}$.
\end{lem*}

\begin{pf}
The fundamental eigenequations for a gPCA of the triplet $(\mtilde
{\locmat},\allowbreak\similar, \wloc)$ are
%
\begin{eqnarray}\label{eq:tcoveigeneq}
\mtilde{\locmat}^T\wloc\mtilde{\locmat}\similar\mat{A}=\mat{A}\bolds{\Psi},
\qquad
\mtilde{\locmat}\similar\mtilde{\locmat}^T\wloc\mat{\matcoltrans}=\mat
{\matcoltrans}\bolds{\Psi},\nonumber
\\[-8pt]
\\[-8pt]
 \eqntext{\mbox{where }
\mat{A}^T\similar\mat{A}=\mat{I}_r,\
\mat{\matcoltrans}^T\wloc\mat{\matcoltrans}=\mat{I}_r,}
\end{eqnarray}
 so that $\locmat\proj[\spweightm]=\mat{\matcoltrans}\bolds{\Psi
}^{1/2}\mat{A}^T$ is the\vadjust{\goodbreak} corresponding gSVD.

Since $\mtilde{\locmat}=\locmat\proj[\spweightm]$, we see that $\mat
{\matcoltrans}$ and $\mat{G}$ from DPCoA are both eigenvectors for the
same matrix, $\locmat\proj[\spweightm]\similar\proj[\spweightm]^T\locmat
^T\wloc$, implying that $\mat{\matcoltrans}$ and~$\mat{G}$ are the
$\wloc$-orthonormal eigenvectors of the same matrix. This implies that
the eigenvalues are the same ($\bolds{\Phi}=\bolds{\Psi}$) and that $\mat
{\matcoltrans}$ and $\mat{G}$ are the same up to a~sign change
(assuming unique eigenvalues).

The resulting coordinates for the locations under DPCoA are given by
$\sitecoord=\dpy\mat{F}=\mat{G}\bolds{\Phi}^{1/2}.$ With gPCA of $(\mtilde
{\locmat},\similar,\wloc)$, the location coordinates are
$\mat{\newX}=\locmat\proj[\spweightm]\similar\mat{A}
=\mat{\matcoltrans}\bolds{\Psi}^{1/2}$
and, therefore, we have that $\sitecoord=\mat{\newX}$---the
coordinates of the locations are the same in the two methods.

The coordinates for the species are given by DPCoA as the rotation of
the coordinates given in $\dpx$ by $\mat{F}$: $\speciescoord=\dpx\mat
{F}$. By the gSVD decomposition of $\dpy$, we can write $\mat{F}^T=\bolds{\Phi}^{-1/2}\mat{G}^T\wloc\dpy$ and,
similarly, $\mat{\matcoltrans}\bolds
{\Psi}^{-1/2}=\locmat\proj[\spweightm]\similar\mat{A}\bolds{\Psi}^{-1}$.\vspace*{1pt}
Remembering that $\dpx\dpx^T=\proj[\spweightm]\dist\proj[\spweightm]^T$,
the final coordinates of the species from DPCoA are given by
\begin{eqnarray*}
\speciescoord&=&\dpx\dpy^T\wloc\mat{G}\bolds{\Phi}^{-1/2}=\dpx\dpx^T\locmat
^T\wloc\mat{G}\bolds{\Phi}^{-1/2}\\
&=&\proj[\spweightm]\dist\proj[\spweightm]^T\locmat^T\wloc\mat{G}\bolds{\Phi
}^{-1/2}\\
&=&\proj[\spweightm]\similar\underbrace{\proj[\spweightm]^T\locmat^T\wloc
\locmat\proj[\spweightm]\similar\mat{A}}_{=\mat{A}\bolds{\Psi}\mathrm{\
from\
{\fontsize{8.36pt}{10pt}\selectfont{\eqref{eq:tcoveigeneq}}}}} \bolds{\Psi}^{-1}
=\proj[\spweightm]\similar\mat{A}
\end{eqnarray*}
 up to the sign change difference between $\mat{G}$ and $\mat
{\matcoltrans}$.
\end{pf}

\section{Kernel analysis and gPCA}\label{app:kernel}
Multivariate kernel methods seek a set of functions $f_1,\ldots,f_k$
from our general data space $\mc{X}$ into $\R{}$, such that the
possible set of functions $f$ form a Reproducing Kernel Hilbert Space
with respect to a kernel function $\mc{K}$ on~$\mc{X}$ [see \citet
{ScholkopfBook} for details]. The solutions for a multivariate Kernel
CCA (or extensions) can be recovered from the eigenequations, assuming
$\kernel[i]$ are invertible
\[
\kernel[{\xi_2}]^{-1}\kernel[2]\kernel[1]\mat{U}_1=\kernel[\xi_1]\mat
{U}_1\bolds{\Lambda}
\]
with the constraint that
\[
\mat{U}_i^T\kernel[{\xi_i}]\mat{U}_i=\bolds{\Gamma}_i
\]
and $\mat{U}_2$ is given by
\[
\mat{U}_2=\kernel[{\xi_2}]^{-1}\kernel[1]\mat{U}_1(\bolds{\Lambda}\bolds
{\Gamma}_1\bolds{\Gamma}_2^{-1})^{-1/2},
\]
where $\kernel[{\xi_i}]=(1-\xi_i)\kernel[i]/n+\xi_i\mat{I},$ and $\bolds{\Gamma}_i$ are diagonals of normalization constants chosen by the
user. Then
the new coordinates of an object $x$ from data set $i$ are given by
$(f_1(x),\ldots,f_k(x))^T=\vc{k}^T\mat{U}_i$, where $\vc[][j]{k}=\mc
{K}_i(x,x_j)$.

Let $\kernel[1]=\colmetric,$ $\kernel[2]=\mat{X}\rowmetric\mat{X}^T,$
and $\xi_1=\xi_2=1$, then we have the eigenequation
\begin{equation}
\mat{X}\rowmetric\mat{X}^T\colmetric\mat{U}_1=\mat{U}_1\bolds{\Lambda}.
\end{equation}
We see that these are equivalent to the gPCA equations, with $\mat
{U_1}=\mat{B}$. Then the coordinates associated with the data matrix
$\mat{X}$ from the kernel method are
\[
\kernel[2]\mat{U}_2=\kernel[2]\kernel[1]\mat{U}_1(\bolds{\Lambda}\bolds
{\Gamma}_1\bolds{\Gamma}_2^{-1})^{-1/2}=\mat{X}\rowmetric\mat
{X}^T\colmetric\mat{U}_1\bolds{\Lambda}^{-1/2}(\bolds{\Gamma}_1\bolds{\Gamma
}_2^{-1})^{-1/2},
\]
while those from the gPCA are
\[
\mat{X}\rowmetric\mat{A}=\mat{X}\rowmetric\mat{X}^T\colmetric\mat{B}\bolds{\Lambda}^{-1/2}.
\]
Choosing the scaling of the eigenvectors so that $(\bolds{\Gamma}_1\bolds{\Gamma}_2^{-1})=\mat{I}$ makes the solutions equivalent.

\section{Inertia and dissimilaries}\label{app:decomposeDiv}
We generalize the results of \citet{Pelissier} to show the derivation
of the dissimilarity and diversity results above.

In gPCA, the term \textit{inertia} is used for the inter-point
similarities, and the total inertia between points is defined as $I(\mat
{X},\rowmetric,\colmetric)=\operatorname{tr}(\colmetric\mat{X}\rowmetric\mat{X}^T)=\sum
\lambda_i.$
Then if $\mat{\newX}_{(r)}$ are the new coordinates of $\mat{X}$
restricted to the first $r$ dimensions and $\mat{\newX}_{(-r)}$ the
remaining $\ncol-r$ dimensions, we can decompose the total inertia into
the inertia of the first $r$ dimensions and that of the remaining $\ncol-r$,
\begin{eqnarray*}
I(\mat{X},\rowmetric,\colmetric)&=&
I\bigl(\mat{\newX}_{(r)},\id[p],\colmetric\bigr)
+
I\bigl(\mat{\newX}_{(-r)},\id[p],\colmetric\bigr)\\
&=&\sum_{i=1}^r \lambda_i +\sum_{i=r+1}^p\lambda_i,
\end{eqnarray*}
and the first $r$ dimensions give maximal possible inertia for $r$ dimensions.

Let $\mat{Y}\in\R{\atotal\times\nspecies}$ be the incidence matrix
for the species variable, where~$\mat{Y}_{(is)}$ is an indicator of the
$i$th observation being species $s$. Let $\mat{Z}\in\R{\atotal\times
\nsites}$ be a similar such incidence matrix for the location variable.
Then the inertia of the eigenanalysis of the triplet $(\mtilde{Y},
\rowmetric[],\mat{D}_\atotal)$, where $\mtilde{Y}$ is the (nonweighted)
centered $\mat{Y}$ and $\mat{D}_\atotal$ is a diagonal matrix of
$\atotal$ elements, will be equal to $H_{\rowmetric[]} (\locbar)$.
Regressing $\mat{Y}$ onto $\mat{Z}$ gives predictions $\mtilde{Y}_{Z}$
and residuals $\mtilde{Y}_{| Z}=\mtilde{Y}-\mtilde{Y}_{Z}$. Then the
total inertia ($I_{\mathrm{T}}$) can be broken into the inertia due to differences
between locations ($I_{\mathrm{B}}$) plus the remaining inertia within
locations~($I_{\mathrm{W}}$),\looseness=1
\[
\operatorname{Inertia}(\mtilde{Y},\rowmetric[],\mat{D}_\atotal)= \operatorname{Inertia}(\mtilde
{Y}_{Z},\rowmetric[],\mat{D}_\atotal) + \operatorname{Inertia}(\mtilde{Y}_{|
Z},\rowmetric[],\mat{D}_\atotal).
\]

Note that $\mtilde{Y}_{Z}=\mat{Z}\tildlocmat$, so that the inertia
explained by $\mat{Z}$ is equal to the inertia of the eigenanalysis of
$\tildlocmat$,
\[
I_{\mathrm{B}}=\operatorname{Inertia}(\mtilde{Y}_Z,\rowmetric[],\mat{D}_\atotal
)=\operatorname{Inertia}(\tildlocmat,\rowmetric[],\wloc).
\]
This implies that the ordination procedures described above best
preserve the between location dissimilarities defined by the metric
$\rowmetric[]$.

\section{The Laplacian and a Laplacian for trees}\label{app:treeLaplacian}

The Laplacian matrix that is associated with the graph is given by $\lap
=\mat{D}-\mat{\adj}$, where $\mat{\adj}$ is the adjacency matrix of the
graph and $\mat{D}$ is the diagonal matrix consisting of the degree of
each vertex. The spectral decomposition of~$\lap$ is closely related to
certain properties of the graph; in particular, there are many results
linking the eigenvalues of $\lap$ with fundamental characteristics of
the graph [see \citet{DiestelBook}]. There are fewer explicit
characterizations of the eigenvectors that hold for all graphs. In a
general way, the eigenvectors corresponding to small eigenvalues of
$\lap$ represent large divisions in the graph (indeed, for $\lambda
_0=0$, we have the eigenvector $\one$ which is an average of all the
nodes); they tend to be zero for large portions of the graph and the
nonzero components are the same sign distinct regions of the graph.
Those eigenvectors corresponding to large eigenvalues tend be dominated
by linear combinations of ``close'' nodes or smaller groups of nodes and
represent the ``noisy,'' small differences within neighboring vertices.
Thus, the eigenvectors of the Laplacian have ``multiscale''
characteristics, particularly those eigenvectors corresponding to the
largest and smallest of the eigenvalues. For data~$\vc{x}$ associated
with a graph, with each element of $\vc{x}$ corresponding to a node in
the graph, the metrics for a graph based on the Laplacian will usually
put greater weight on the eigenvectors corresponding to small
eigenvalues, for example, $1/\lambda_i$ or $\exp(-1/\lambda_i)$. This
choice corresponds to the behavior of the eigenvectors.

The Laplacian gives the covariance between nodes from a useful model
for describing relationships among the nodes---a model of diffusion of
information through the graph. The covariance from this model is given
by $\exp(-2\alpha\lap)$, known as the \textit{heat kernel} of the graph
[see \citet{kondor2002} for review]. Of course, this is equivalent to
weighting the eigenvectors of the Laplacian with weight function $\exp
(-\alpha\lambda_i)$.

A phylogenetic tree is, of course, a graph, and the Laplacian of a tree
and the distances between nodes on a tree are quite simply related
[\citet{Bapat2005}].
Let $\dist_T$ be the distance matrix of the patristic distances between
\textit{all} the nodes of the tree (internal nodes as well as the
leaves), and let $\mat{L}$ be the Laplacian of the tree with weights
$1/d(r,s)$ on each edge. Then we have that
\[
\mat{L}=\vc{v}\vc{v}^T/ \sum d(r,s) -2 \dist_T^{-1},
\]
where for a phylogenetic tree $\vc{v}$ is $-1$ or $1$ depending on
whether the node is a leaf of the tree or not.

However, since our data is observed on only certain nodes of the graph---the
leaves of the tree---we need a metric that gives a relationship only
between the leaves. If we use the Laplacian as our phylogenetic metric,
we would have to constrain ourselves to the portion of the metric that
corresponds to the relationships between just the leaves, $\mat
{L}_\nspecies$. If we took as our metric the inverse of the Laplacian---which corresponds
to an appropriate ordering of the eigenvectors\vspace*{1pt} by
weighting each by $1/\lambda_i$---we have that $\mat{L}^{-1}_\nspecies
$ is given by
\[
\mat{L}^{-1}_\nspecies=c \vc{\bolds\gamma}\vc{\bolds\gamma}^T -1/2 \dist_\nspecies
,  \qquad
\mbox{where }
c=(8 \one^T\dist_{\nspecies\times I}\one)^{-1},\
\vc{\bolds\gamma}=\dist_T\vc{v},
\]
and $\dist_\nspecies\in\R{\nspecies\times\nspecies}$ is the distance
matrix restricted to the distances between leaves of the tree and $\dist
_{\nspecies I}\in\R{\nspecies\times\nspecies-1}$ is the distance matrix
restricted to the distances between the leaves of the tree and
$\nspecies-1$ internal nodes of the tree. This is an expression
somewhat similar to our similarity matrix for DPCoA, but note that a
gPCA based on $\mat{L}^{-1}_S$ is not equivalent to DPCoA because $\proj
\vc{\bolds\gamma}\vc{\bolds\gamma}\proj^T$ does not vanish.

However, restricting the metric to those portions dealing only with the
leaves makes the metric difficult to interpret. The Laplacian
restricted to the leaves will no longer have the same eigenvectors as
the Laplacian and thus loses its connection to the behavior shown by
the eigenvectors of the Laplacian. Furthermore, from the point of view
of covariance modeling, the phylogenetic tree represents an
evolutionary story that is more directly modeled by $\tcov$.

\section{\texorpdfstring{Eigenvectors of $\tcov$ for a phylogenetic tree}
{Eigenvectors of Sigma for a phylogenetic tree}}\label{app:eigenvectors}

Note the block structure in $\tcov$: if the root ancestor, $\troot$,
has immediate descendants $\mc{P}_1$ and $\mc{P}_2$, then the
covariance between any of the existing descendants of $\mc{P}_1$ and
those of $\mc{P}_2$ will be $0$. Thus, we can order the rows and
columns of $\tcov$ so that
%
\begin{equation}\label{eq:blockCov}
\tcov=
\pmatrix{
\tcov_1 & \myvec{\bm\varnothing} \cr
\myvec{\bm\varnothing} & \tcov_2
}
,
\end{equation}
where $\tcov_1$ is a $\nspecies_1\times\nspecies_1$ matrix, $\nspecies
_1$ is the number of extant species descended from $\mc{P}_1$, and
similarly with $\tcov_2$. This means that the eigenvectors of $\tcov$
must be of the form
%
\begin{equation}\label{eq:blockeigvec}
\pmatrix{
\myvec[1i]{v} \cr
\myvec{\bm\varnothing}
}
  \quad \mbox{or} \quad
\pmatrix{
\myvec{\bm\varnothing} \cr
\myvec[2j]{v}
}
,
\end{equation}
where $\{\myvec[1i]{v}\}_{i=1}^{\nspecies_1}$ are the eigenvectors of
$\tcov_1$ and $\{\myvec[2j]{v}\}_{j=1}^{\nspecies_2}$ are the
eigenvectors of $\tcov_2$. Therefore, every eigenvector of $\tcov$, at
a minimum, must be only nonzero for one of the lineages.

Indeed, if we think back to the definition of $\tcov$, the elements of
the blocks~$\tcov_1,\tcov_2$ are themselves rank-1 perturbations of
block diagonal matrices:
\begin{equation}
\tcov_1=
\pmatrix{
\tcov_{11} & \myvec{\bm\varnothing} \cr
\myvec{\bm\varnothing} & \tcov_{12}
}
+c_1\one\one^T, \qquad \tcov_2=
\pmatrix{
\tcov_{21} & \myvec{\bm\varnothing} \cr
\myvec{\bm\varnothing} & \tcov_{22}
}
+c_2\one\one^T,
\end{equation}
 where $c_1=\treedist{\troot}{\mc{P}_1}$ and $c_2=\treedist
{\troot}{\mc{P}_2}$ (here we have assumed that $\vc{t}\propto\one$ for
simplicity). This same logic continues so that each sub-block can be
written as a block matrix plus a rank-one perturbation.
$\tcov$ thus consists of such nested rank-1 perturbations of block matrices.

The claims in the literature for a relationship of the eigenvectors of
$\tcov$ to the partitions of the tree all stem from the comments of
\citet{cavalli1975}. They make assertions which they prove only in the
case of a tree with four leaves ($\nspecies=4$) and
under the assumption of a constant rate of evolution ($\vc{t}\propto\one
$). One assertion is true: for any terminal bifurcation node (a node
whose two descendants are existing species or leaves of the tree),
there is an eigenvector of $\tcov$ that has elements that are positive
for one of the species, negative for the other and zero for all other species.
In addition, we see that because of the block structure, every
eigenvector of $\tcov$, at a minimum, must consist of zero elements for
one branch of the tree.

Beyond this, \citet{cavalli1975} describe ``usual'' behavior of the
eigenvectors, but their ideas do not scale as the size of the tree
increases. The nested block structure of $\tcov$ still has the effect
of creating eigenvectors with some structure to them, though not as
easily classified as suggested in \citet{cavalli1975}. Generally the
structure of the eigenvectors will not be directly related to a
partition in the tree. In practice, the eigenvectors often have some
relation to the bifurcations of the tree, particularly the deeper
(earlier in time) bifurcations and of course the terminal bifurcations.
The other eigenvectors often have clumps of positive and negative
elements that correspond to subtrees of the tree, and we often
empirically see as the eigenvalues get smaller some sort of
concentration of large values in only a few species.

\section{Ellipses in DPCoA plots}\label{app:ellipses}
The ellipse plots given in Figure 2(b) are provided by the \texttt{ade4}
package and represent a location vector, $\vc[\ell]{x}$, as an ellipse.
For completeness, we explain here what \texttt{ade4} is plotting.

Let the \textit{species} coordinates as transformed into two dimensions
by the ordination of the \textit{locations} be given in the columns of
$\speciescoord_2\in\R{\nspecies\times2}$. Then the ellipsoid for $\vc
[\ell]{x}\in\R{\nspecies}$ is defined by
\[
\vc{v}^T(\speciescoord_2^T\mat{D}_{\locprofile{l}}\speciescoord
_2)^{-1}\vc{v}=1,
\]
where the ellipse is centered at $\locprofile{l}$.\vadjust{\goodbreak}

This curve consists of the points with norm $1$ in the Mahalanobis
metric, only the estimate of the variance in Mahalanobis distance is
calculated with weights on the points (species) given by $\locprofile
{l}$. Equivalently, the ellipses in Figure 2(b) will have major and minor
axes in the direction of the weighted principal components of the
coordinates of the species in $\speciescoord_2$, with the lengths of
the axes given by the weighted standard deviation of the species
coordinates in those directions (an ellipse defined by the equation $\vc
{x}^T\mat{Q}\vc{x}=1$ will have major and minor axes in the directions
of the eigenvectors of $\mat{Q}$ with lengths given by $1/\sqrt{\lambda
_i}$, where $\lambda_i$ is an eigenvalue of $\mat{Q}$).
\end{appendix}



\section*{Acknowledgments}
The author would like to thank Susan Holmes for many helpful
conversations and reviews of previous drafts, and Elisabeth Bik, Les
Dethlefsen, Paul Eckburg and David Relman for discussions regarding the
data and data analysis and for use of the data.

\printaddresses

\end{document}